\documentclass[traditabstract]{aa} % for the abstract without structuration
\usepackage{graphicx}
\usepackage{psfig}
\usepackage{txfonts}
\usepackage{natbib}
\usepackage{longtable}
\usepackage{color}
\begin{document}
\newcommand{\zabs}{\ensuremath{z_{\rm abs}}}
\newcommand{\avg}[1]{\left< #1 \right>} % for average
\newcommand{\PN}{\color{red} PN:~}

\title{Neutral atomic-carbon QSO absorption-line systems at $z>1.5$:}
%Cold gas at high redshift: Dust reddening and obscuration of
%SDSS quasars induced by the strongest known neutral-carbon
%absorbers

\subtitle{Sample selection, \ion{H}{i} content, reddening, and 2175~\AA\ extinction
 feature\thanks{Based on data from the Sloan Digital Sky Survey (SDSS) and dedicated
 follow-up observations carried out at the European Southern Observatory (ESO) under
 programmes 082.A-0544 and 083.A-0454 (P.I. Ledoux) using the Ultra-violet and
 Visual Echelle Spectrograph (UVES) installed at the Nasmyth-B focus of the Very
 Large Telescope (VLT), Unit-2 (Kueyen), on Cerro Paranal, Chile.}}

\author{C. Ledoux\inst{1}
   \and P. Noterdaeme\inst{2}
   \and P. Petitjean\inst{2}
   \and R. Srianand\inst{3}
}

%\offprints{cledoux@eso.org}

\institute{European Southern Observatory, Alonso de C\'ordova 3107,
  Casilla 19001, Vitacura, Santiago 19, Chile\\
  \email{cledoux@eso.org}
  \and Institut d'Astrophysique de Paris, CNRS and UPMC Paris 6, UMR
  7095, 98bis boulevard Arago, 75014 Paris, France
  \and Inter-University Centre for Astronomy and Astrophysics, Post
  Bag 4, Ganeshkhind, 411\,007 Pune, India }

\date{Accepted for publication 20 April 2015}

\authorrunning{C. Ledoux et al.}
\titlerunning{\ion{C}{i}-selected QSO absorption-line systems at $z\sim 2$
%The strongest \ion{C}{i}-absorber population.
%Cold gas at high redshift.
}

\abstract{
  We present the results of a search for cold gas at high redshift along
  QSO lines-of-sight carried out without any a priori assumption on the neutral
  atomic-hydrogen content of the absorbers. To do this, we systematically looked
  for neutral-carbon (\ion{C}{i}) $\lambda\lambda$1560,1656 transition lines in
  41\,696 low-resolution QSO spectra ($1.5<z_\mathrm{em}<4.46$) from the SDSS-II
  - Data Release seven - database. \ion{C}{i} absorption lines should indeed probe the
  shielded gas in the neutral interstellar medium of galaxies more efficiently than
  traditional tracers such as neutral atomic-hydrogen (\ion{H}{i}) Damped
  Lyman-$\alpha$ (DLA) and/or \ion{Mg}{ii} systems.\\
  We built up a sample of 66 \ion{C}{i} absorbers with redshifts in the
  range $1.5<z<3.1$ and rest-frame equivalent
  widths $0.1<W_\mathrm{r}(\lambda 1560)<1.7$~\AA. The completeness limit
  of our survey is $W_\mathrm{r,lim}(\lambda 1560)\simeq 0.4$~\AA. \ion{C}{i}
  systems stronger than that are more than one hundred-times rarer than DLAs at
  $z_\mathrm{abs}=2.5$. The number of \ion{C}{i} systems per unit redshift
  is found to increase significantly below $z=2$. We suggest that the \ion{C}{i} absorbers are
  closely related to the process of star formation and the production of dust in
  galaxies and that their cosmic evolution is driven by the interplay between
  dust shielding and the evolution of the ultra-violet background at
  $\sim 10$~eV.\\
  We derive the neutral atomic-hydrogen content of the \ion{C}{i} systems
  observable from the southern hemisphere from VLT/UVES spectroscopy and find
  that a majority of them are sub-DLAs with $N(\ion{H}{i})\sim 10^{20}$ atoms~cm$^{-2}$.
  The dust content of these absorbers is yet significant as seen from the redder
  optical colours of the corresponding background QSOs and their reddened spectral energy distributions,
  with $E($B-V$)$ values up to $\sim 0.3$. The overall $N(\ion{H}{i})$ distribution of
  \ion{C}{i} systems is relatively flat however. As a consequence, among the
  \ion{C}{i} systems classifying as DLAs there is a probable excess of strong DLAs
  with $\log N(\ion{H}{i})>21$ (atoms~cm$^{-2}$) compared to systematic DLA
  surveys. While the dust content of such systems is significant, their dust-to-gas
  ratio must still be limited. Indeed, strong DLAs having large amounts of shielded gas and
  dust producing stronger reddening and extinction of the background QSOs if they
  exist should have been missed in the current magnitude-limited QSO sample.\\
%Krumholz et al. 2009
  We study empirical relations between $W_\mathrm{r}($\ion{C}{i}$)$, $N(\ion{H}{i})$,
  $E($B-V$)$ and the strength of the 2175~\AA\ extinction feature, the latter
  being detected in about 30\% of the \ion{C}{i} absorbers. We show that
  the 2175~\AA\ feature is weak compared to Galactic lines-of-sight exhibiting the
  same amount of reddening. This is probably the consequence of current or past
  star formation in the vicinity of the \ion{C}{i} systems. We also find that the
  strongest \ion{C}{i} systems tend to have the largest amounts of dust and that
  the metallicity of the gas and its molecular fraction is likely to be high in a large number
  of cases. The \ion{C}{i}-absorber sample presented here hence provides ideal
  targets for detailed studies of the dust composition and molecular species at high redshift.}


\maketitle

%------------------------------------------------------------------------------

\section{Introduction}

\label{sec:intro}

Understanding the mechanisms of star formation at high redshifts is
central to our knowledge of how galaxies formed and subsequently
evolved chemically. This is specially true at $z\sim 2$ when the
cosmic star-formation activity was highest \citep{2014ARA&A..52..415M}.
Stars form out of cold gas, metals and dust in molecular clouds
\citep[e.g.][]{2006ARA&A..44..367S} in the interstellar medium (ISM)
of galaxies. In turn, the
radiative and mechanical feedbacks from stars have a strong impact on the
physical state of the ISM. Studying the ISM at high redshifts, and in particular
deriving the physical properties of the diffuse molecular phase in galaxies, is
therefore crucial for understanding how stars formed in the early Universe.

The best way to derive physical properties accurately is to detect the
tracers of the cold gas in absorption \citep[see][]{2014A&A...566A.112M}.
The neutral, shielded and possibly cold gas clouds at high redshifts can
be searched for in the radio domain by targeting the
neutral atomic-hydrogen (\ion{H}{i}) 21-cm absorption
line \citep[e.g.,][]{2009MNRAS.398..201G}. However, systematic - blind -
surveys have to await the increased sensitivity of new facilities such as
MeerKAT/SKA and ASKAP \citep{2009arXiv0910.2935B,2012MNRAS.426.3385D}.
On the other hand, such a gas can be efficiently traced in the optical wave-bands by
detecting the redshifted \ion{H}{i} damped Lyman-$\alpha$
\citep{2005ApJ...635..123P,2009A&A...505.1087N,2012A&A...547L...1N}
and/or strong \ion{Mg}{ii}
\citep[e.g.,][]{2011AJ....141..137Q,2011MNRAS.416.1871B} lines
imprinted in the spectra of bright enough background sources such as QSOs
or the rapidly fading $\gamma$-ray burst (GRB) afterglows \citep[for
the latter, see][and references therein]{2009ApJS..185..526F}.

Damped Lyman-$\alpha$ systems (hereafter DLAs) observed in QSO spectra
have column densities of $N(\ion{H}{i})\ge 2\times
10^{20}$ atoms cm$^{-2}$ and are known to contain most of the neutral gas
in the Universe in the redshift range $0<z<5$ \citep[see][for a
review]{2005ARA&A..43..861W}. It has been shown however that DLAs
typically probe warm ($T\ga 3000$~K) and diffuse
($n_\mathrm{H}<1$~cm$^{-3}$) neutral gas
\citep[e.g.,][]{2000A&A...364L..26P,2012MNRAS.421..651S}. The
metallicity of DLAs is generically low, i.e., on an average about
1/30$^{\rm th}$ of Solar
\citep{2006fdg..conf..319P,2012ApJ...755...89R} and their dust-to-gas
ratio is typically less than one-tenth of what is observed in the Galactic ISM
\citep[e.g.,][]{2008A&A...478..701V}. This probably explains the low
detection rates of molecular hydrogen
(H$_2$) in DLAs where only about 10\% of the QSO lines-of-sight
intercept H$_2$-bearing gas down to a limit of $N($H$_2)\sim 10^{14}$ molecules
cm$^{-2}$ (e.g., \citealt{2008A&A...481..327N,2014arXiv1402.2672B};
for searches for H$_2$ in DLAs originating from the host galaxies of
GRBs, see \citealt{2009A&A...506..661L,2013A&A...557A..18K}).

Based on the observed correlation between metallicity and dust depletion
in DLAs \citep{2003MNRAS.346..209L}, DLAs with high metallicity are
expected to contain more dust and therefore to exhibit larger H$_2$
fractions \citep[see][]{2006A&A...456L...9P}. However, even in
DLAs with the highest metallicities typical dust signatures like
reddening of the background QSOs, the 2175~\AA\ extinction feature
(hereafter also called ultra-violet [UV] bump), or diffuse interstellar bands, are
not apparent. Even in the rare cases with H$_2$ detections, the inferred
molecular fractions are low and typical of what is seen in Galactic diffuse atomic gas
with $f($H$_2)\equiv 2N($H$_2)/[2N($H$_2)+N(\ion{H}{i})]\la 0.01$ and often
much lower than this \citep[see][]{2008A&A...481..327N}. The primary
reason for this is that the cold and dusty phases are missed probably
because of their reduced cross-sections relative to that of the more
pervasive warm neutral ISM \citep{2006ApJ...643..675Z}. Direct
evidence for the relatively small physical sizes ($\la$~0.15~pc) of
H$_2$-detected clouds in DLAs recently came from the observation of partial
coverage of the QSO broad-line emitting region
(\citealt{2011MNRAS.418..357B,2015MNRAS.448..280K}).

H$_2$-detected clouds in DLAs are found to have kinetic
temperatures in the range $T\sim 70-200$~K and particle densities
$n_\mathrm{H}\sim 1-100$~cm$^{-3}$
\citep[e.g.,][]{2005MNRAS.362..549S}. When detected, H$_2$ is usually
coincident with neutral atomic carbon \citep[\ion{C}{i}; see
also][]{1999ASPC..156..121G}. This is due to the fact that the
ionization potential of neutral carbon ($11.26$~eV) is similar
to the average energy of the Lyman-Werner photons that dissociate H$_2$.
Therefore, shielding of UV photons is essential for these species to remain
at detectable levels. Carbon
monoxide (CO) has long escaped detection even in DLAs with detected
H$_2$, down to $N($CO$)\sim 10^{12}$ molecules cm$^{-2}$ \citep[see,
e.g.,][]{2002MNRAS.332..383P}. This is not surprising since CO, with a
dissociation energy of $11.09$~eV, needs to be even more shielded than
H$_2$ and \ion{C}{i} to be detected. After CO UV absorption bands were detected
for the first time at high redshift, in a sub-DLA towards the QSO
SDSS~J\,143912.05$+$111740.6 \citep{2008A&A...482L..39S}, it became
clear that the best place to detect CO in absorption at high redshift
are the systems with strong \ion{C}{i} absorption. Following this strategy
allowed us to detect carbon monoxide subsequently in five additional systems
\citep{2009A&A...503..765N,2010A&A...523A..80N,2011A&A...526L...7N}.

In Galactic translucent interstellar clouds, CO starts to be produced
in significant amounts when neutral atomic carbon becomes the dominant
carbon species and a large fraction of hydrogen turns molecular
\citep{2006ARA&A..44..367S}. The strength of the \ion{C}{i} absorption
is expected to be such that it could be detected even in a low resolution
spectrum. We therefore embarked in a systematic search for
\ion{C}{i} absorption in QSO spectra from the SDSS-II - Data Release seven
(hereafter DR\,7) - database. In this paper, we present the results of this
search and the basic properties of the detected \ion{C}{i} absorbers.
% A companion paper will present the systematic search for CO in this
% sample.
Note that here we will equally refer to QSO absorption-line systems detected
through \ion{C}{i} absorption as ``\ion{C}{i} systems'' or ``\ion{C}{i} absorbers''.

In Sect.~\ref{sec:identification}, we describe our selection and identification of
\ion{C}{i} absorbers. We discuss the properties of the sample in terms of intervening
\ion{C}{i}-absorber number per unit redshift, proximate systems, and \ion{C}{i}
rest-frame equivalent widths, in Sects.~\ref{sec:nz}, \ref{sec:prox} and \ref{sec:W},
respectively. We then assess the impact of the \ion{C}{i} absorbers on their respective
background QSOs both from the observed QSO optical
colours (Sect.~\ref{sec:colours}) and the reddening these systems induce on the QSO
spectral energy distributions (Sect.~\ref{sec:ebv}). In Sect.~\ref{sec:nhi}, we present
the \ion{H}{i} column-density distribution of the \ion{C}{i} systems
from VLT/UVES spectroscopy. In Sect.~\ref{sec:relations}, we investigate
empirical relations between neutral atomic-carbon and neutral atomic-hydrogen
contents, QSO reddening and the strength of possible 2175~\AA\ extinction features
(whose measurements are described in Sect.~\ref{sec:abump}). We summarise our
findings and conclude in Sect.~\ref{sec:conclusions}.

Throughout this paper, we assume a standard $\Lambda$ cold dark-matter cosmology
with $H_0=70$~km\,s$^{-1}$\,Mpc$^{-1}$, $\Omega_\Lambda=0.7$
and $\Omega_\mathrm{M}=0.3$.

\section{\ion{C}{i} absorption-line selection and identification}

\label{sec:identification}

We systematically searched for \ion{C}{i} absorption lines in high-redshift QSO
spectra from the Sloan Digital Sky Survey \citep{2000AJ....120.1579Y} -- DR\,7 \citep{2009ApJS..182..543A} -- quasar catalogue
\citep{2010AJ....139.2360S}. This survey imposed an i-band magnitude cut
of 19.1 for QSO candidates whose colours indicate a probable redshift smaller
than $\sim 3$. The spectra cover the wavelength range 3800--9200~\AA\ at
a resolving power $R\sim 2000$.

\begin{figure}[htb]
\resizebox{\hsize}{!}{\includegraphics{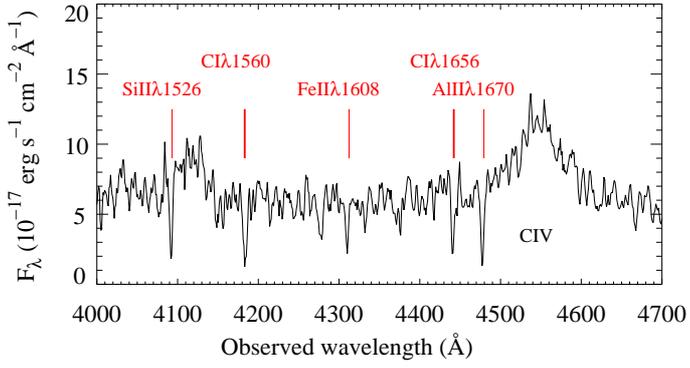}}
\caption{SDSS spectrum of the $z_\mathrm{em}=1.94$ QSO
  J\,0815$+$2640. The \ion{C}{i}\,$\lambda\lambda$1560,1656
  absorption lines detected at $z=1.681$ are indicated along with several
  low-ionization metal lines at the same redshift. Among the latter,
  the simultaneous presence of \ion{Si}{ii}\,$\lambda$1526
  and \ion{Al}{ii}\,$\lambda$1670 lines was required by our detection
  algorithm to minimize the probability of chance
  coincidences. In this system, \ion{Fe}{ii}\,$\lambda$1608 absorption
  is also observed even though relatively weak.\label{fig:discovery}}
\end{figure}

We implemented a dedicated IDL procedure to detect and identify absorption-line
features in SDSS QSO spectra automatically. Since SDSS spectra are log-lambda
binned, the pixels have constant velocity size ($\approx 69$~km\,s$^{-1}$).
This makes it straightforward to cross-correlate the spectra with an emission- or
absorption-line template. We
used the method introduced by \citet{2010MNRAS.403..906N} to search for
[\ion{O}{iii}]\,$\lambda\lambda$4959,5007 emission and \ion{Mg}{ii}
absorption lines. We first normalized the spectra iteratively using Savitzky-Golay filtering.
This consists of smoothing the spectra by convolving them with
a Savitsky-Golay kernel that preserves the sharp QSO emission-line peaks but ignores
narrow features such as metal absorption lines, bad CCD pixels and sky emission-line
residuals. Deviant pixels and their neighbours are then masked out and the
resulting data is convolved again in the same way, and so on and so forth. This procedure has the
major advantage as no {\sl a priori} assumption is required about the functional
form of the QSO continuum (i.e., power law or other) and in addition it is computationally
extremely fast.
We then cross-correlated the normalised spectra with a synthetic profile
of \ion{C}{i}\,$\lambda\lambda$1560,1656 absorption lines. We looked for the positive
correlation signal together with peak absorptions detected at more than $2\sigma$
and $2.5\sigma$, respectively, and differing by less than a factor of three.
%{\bf Note that only this last secondary criterium the correlation criterium itself is not sensitive on the continuum}
The simultaneous detections of the \ion{Si}{ii}\,$\lambda$1526
and \ion{Al}{ii}\,$\lambda$1670 absorption lines were required to support the
identifications of the two features as \ion{C}{i} and hence minimize the probability of
chance coincidence. Spurious detections ($\sim 50$\%), most
of them close to the detection limit, were identified visually and removed from
the sample. In total, we find 66 systems, one of which is shown in
Fig.~\ref{fig:discovery}.

The search for \ion{C}{i} lines was limited to the regions of the spectra
redwards of the QSO Lyman-$\alpha$ emission line to avoid the spurious coincidences
that are frequent in the
Lyman-$\alpha$ forest. The wavelength range above 7200~\AA\ was also not considered
in the search to avoid regions of the spectra heavily affected by residuals resulting
from the sky emission-line subtraction.
We requested that the search window encompasses the wavelengths of
the \ion{Si}{ii}\,$\lambda$1526 and \ion{Al}{ii}\,$\lambda$1670 absorption lines
of the putative \ion{C}{i} systems so that the validity of a system does not rely
solely on the detection of two transitions (see above). For a given line-of-sight, the
redshift lower bound ($z_\mathrm{min}$) of the \ion{C}{i} search is therefore
the largest value between $z=3820/1526-1\simeq 1.50$ and
$(1+z_\mathrm{em})\times 1215/1526-1\simeq 0.8\times z_\mathrm{em}-0.2$,
where $z_\mathrm{em}$ is the QSO emission redshift. The redshift upper bound
($z_\mathrm{max}$) of the search is the smallest value between
$z_\mathrm{em}+0.1$ (to not exclude a priori proximate systems with infalling velocities
of up to $+5000$~km\,s$^{-1}$) and $z=7200/1656-1\simeq 3.35$. In order to avoid
too many false positives at low signal-to-noise (S/N) ratio, we requested the median
S/N ratio per pixel to be larger than four for a given spectrum to be actually scanned.
This resulted in a sample of 41\,696 QSOs with $1.5<z_\mathrm{em}<4.46$ whose spectra
were searched for intervening or proximate \ion{C}{i} absorbers. Note that we did not
initially reject Broad Absorption-Line (BAL) quasars because our procedure follows the QSO continuum locally
and can detect narrow absorption lines embedded in broad and not-fully saturated troughs.
Regions of deep absorption are de facto avoided when we study the number of
intervening \ion{C}{i} absorbers per unit redshift in Sect.~\ref{sec:nz} as they have
low S/N ratio per pixel.
%The reason is that the 
%procedure based on Savitsky-Golay filter determines a local continuum. 
%Therefore regions of strong and broad absorptions are avoided
%because of the low S/N ratio at the bottom of the lines.}

% Additionally, the present \ion{C}{i} survey is $\sim 50$\% complete
% for the typical QSO spectrum in our sample (with median S/N ratio
% per pixel of 10) corresponding to $W_\mathrm{r,lim}(\lambda
% 1560)\approx 0.29$~\AA.

Table~\ref{tab:ci} summarises our \ion{C}{i} sample which we refer to
in the following as the overall sample. QSO names with J\,2000
coordinates are given for each absorber. No line-of-sight is found to
feature more than one system\footnote{Note however that there is a second \ion{C}{i} 
system towards SDSS~J\,234023.67$-$005327.1, which, with
$z_\mathrm{abs}=1.36$, falls below the redshift cut-off of our survey. This
system happens to be detected in 21-cm absorption hence is also related to cold gas
\citep[see][]{2009MNRAS.398..201G,2010ApJ...712L.148K}.}. 
The SDSS plate and fibre numbers as well as the MJD are also provided in
the table as useful
cross-references. The QSO emission redshifts derived by the SDSS team
are indicated together with the absorption redshifts and rest-frame equivalent
widths of the \ion{C}{i} lines. The latter were carefully determined by us
for each individual system. In the next column of the table, we specify the
average S/N ratios per pixel in the regions the two \ion{C}{i} lines are
located. At $z_\mathrm{abs}>2.2$, the Lyman-$\alpha$ line of the
systems is also covered by the SDSS spectra. We therefore provide in
the table a determination of the total neutral atomic-hydrogen column
density of these systems following the method developed by
\citet{2009A&A...505.1087N}. The reliability of the latter is
confirmed in a number of cases with follow-up high-resolution VLT/UVES spectroscopy
(see the last column of Table~\ref{tab:ci}, and Sect.~\ref{sec:nhi}).
This column also has additional $N($\ion{H}{i}$)$ measurements
for several low-redshift systems for which Lyman-$\alpha$ absorption
is not covered in the SDSS spectrum.

\section{Sample properties}

It must be noted that in this survey \ion{C}{i} systems are found without
any presumptions on the presence of neutral atomic hydrogen, i.e., the \ion{C}{i} systems
found in this work do not necessarily have to be DLAs. Moreover, DLA absorbers
can be observed in SDSS spectra only when their redshifts are larger than 2.2 while
\ion{C}{i} lines can be identified down to $z_\mathrm{abs}\simeq 1.50$.
As stated before, the reality of the identified \ion{C}{i} systems in our sample was
checked by visual inspection. Therefore, we believe the \ion{C}{i} detections are
secure. We discuss the completeness of the survey in Sect.~\ref{sec:comp}.

\subsection{Line equivalent widths}

\label{sec:W}

\begin{figure}[htb]
%\centering{\hbox{
%\psfig{figure=CIEW.ps,width=8.9cm,clip=,bbllx=25.pt,bblly=115.pt,bburx=543.pt,bbury=704.pt,angle=90.}}}
\resizebox{\hsize}{!}{\includegraphics[angle=90.]{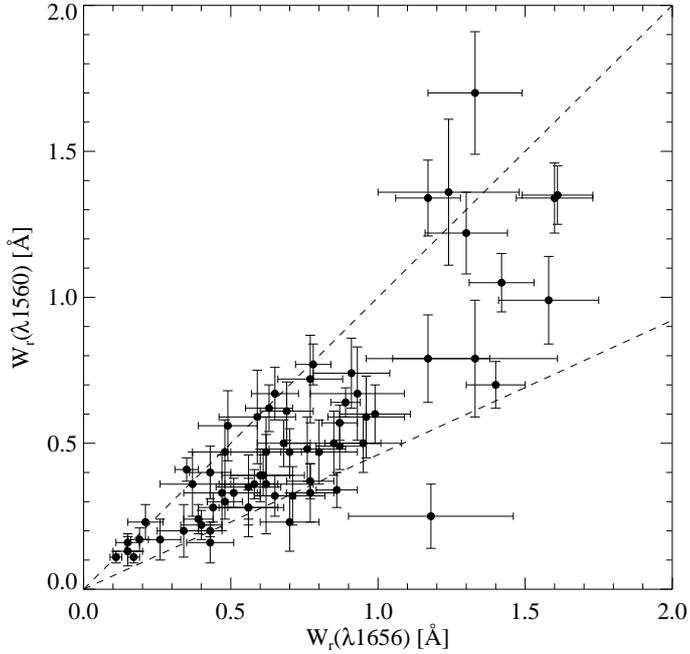}}
\caption{Rest-frame equivalent widths of \ion{C}{i} absorption lines
  measured in SDSS spectra and associated $1\sigma$ uncertainties. The
  upper dashed line shows the 1:1 relation expected in case of heavily saturated
  profiles. The lower dashed line shows the expectation for absorption lines on
  the linear part of the curve-of-growth.
%  {\bf The equivalent widths above which the survey is complete are
%  indicated by dotted lines.}
  \label{fig:ew}}
\end{figure}

Because we will rely on them in the analysis, we here seek to verify the robustness and accuracy of the equivalent-width measurements of \ion{C}{i}-absorption lines performed in SDSS spectra. For this purpose, we plot in Fig.~\ref{fig:ew} the measured rest-frame equivalent widths of the \ion{C}{i}\,$\lambda\lambda 1560,1656$ lines versus each other. It appears that except in one case the strengths of the two lines are in the expected
range. This gives confidence in the derived values and their associated uncertainties. In the case of the outlier seen in the lower part of the plot (i.e., at
$z_\mathrm{abs} =1.526$ towards SDSS~J\,125552.60$+$223424.4), an unrelated blend to
the $\lambda$1656 line is a probable reason for the observed deviation.

All the systems are located within about $2\sigma$ of the boundaries defined by the optically-thin regime on one hand and the relation expected for heavily saturated profiles on the other hand. We note that because of their large equivalent widths ($W_\mathrm{r}\ga 0.4$~\AA) most of the absorbers, especially those with equivalent-width ratios consistent with the optically-thin regime, are probably made of numerous velocity components.

Here, we assumed that the \ion{C}{i} ground-state is solely responsible for the absorption lines while in reality the absorption from the two fine-structure energy levels of the neutral-carbon ground state ($^3$P$_1$ and $^3$P$_2$) could in principle contribute mildly to the measured equivalent widths. However, this will affect the equivalent widths of both of the $\lambda 1560$ and $\lambda 1656$ lines in the same way so that any departure from the assumed relations due to this blending will be small.

\subsection{Completeness}

\label{sec:comp}

Before discussing the number of \ion{C}{i} absorbers per unit redshift ($n_\mathrm{\ion{C}{i}}$;
see the following section), we first need to estimate the completeness of the sample. Given the resolving
power $R$ of the SDSS spectra, the \ion{C}{i}\,$\lambda 1560$ line rest-frame equivalent width limit is
given by:
\begin{equation}\label{eq:ewl}
  W_\mathrm{r,lim}(\lambda 1560)\simeq n\times\frac{1560}{R}\times\mathrm{S/N}^{-1}
\end{equation}
where $n=2$ is the number of standard deviations above which the peak absorption must be
detected and $\mathrm{S/N}>4$ is the limit on the signal-to-noise ratio per pixel at the
corresponding line position. Note that the FWHM of the lines is sampled by two velocity pixels of
constant value. Our survey should therefore be complete down to
$W_\mathrm{r,lim}(\lambda 1560)\simeq 0.4$~\AA.

We checked the exact level of completeness of our survey at this equivalent-width limit by
implementing the following procedure. For this purpose, we used the same data set that
we used for the calculation of
$n_\mathrm{\ion{C}{i}}$ in Sect.~\ref{sec:nz}, i.e., the same quasar sample,
the same [$z_\mathrm{min}$,$z_\mathrm{max}$] values and the same mean $\mathrm{S/N}>4$
limit. We then randomly selected 1000 QSO spectra and introduced an artificial \ion{C}{i} system
of rest-frame equivalent width $W_\mathrm{r}(\lambda 1560)$ at different positions in
the spectra where the local S/N ratio at both \ion{C}{i} lines is larger than four. The
distribution of the equivalent width ratio of the $\lambda$1560 and $\lambda$1656 lines is assumed to
be a normal distribution with a dispersion corresponding to what is seen in Fig.~\ref{fig:ew}.
Note however that neither the equivalent width ratio nor the exact number of artificial systems
used in the simulation has any significant impact on the completeness we infer.

\begin{figure}[htb]
%\centering{\hbox{\hspace{0.3cm}
%\psfig{figure=CIcom.eps,width=8.2cm}}}
\resizebox{\hsize}{!}{\includegraphics{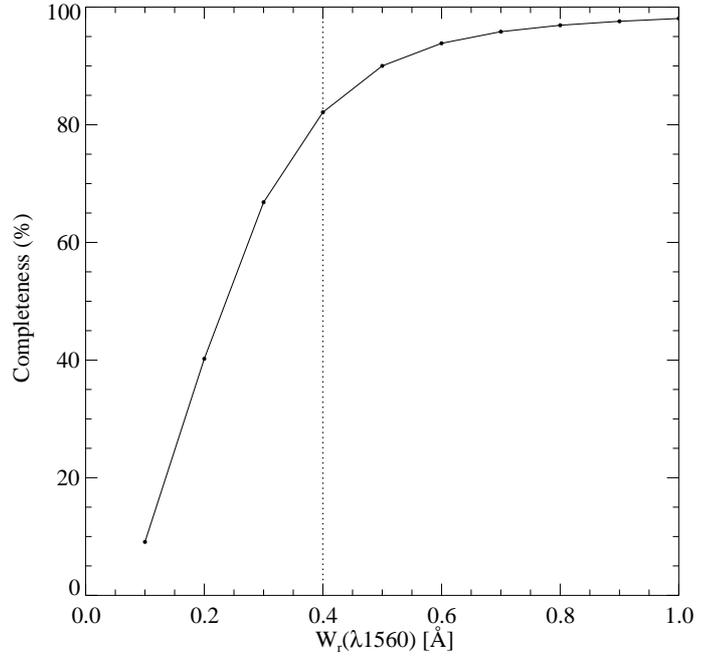}}
\caption{Completeness of our \ion{C}{i} search as a function of
the \ion{C}{i}\,$\lambda$1560 rest-frame equivalent width. Each point is based
on a total of 40\,000 artificial \ion{C}{i} systems introduced in 1000 randomly-selected
SDSS QSO spectra. The vertical dotted line shows our chosen completeness limit
of $W_\mathrm{r}(\lambda 1560)=0.4$~\AA.\label{fig:CIcomp}}
\end{figure}

We implemented about 40\,000 \ion{C}{i} systems that we sought to recover by using
the same automatic procedure described in Sect.~\ref{sec:identification}. We
varied $W_\mathrm{r}$ over the range 0.1-1.0~\AA\ and defined the completeness as the ratio
of the number of recovered systems to the total number of systems introduced in the spectra.
The results are displayed in Fig.~\ref{fig:CIcomp}. It can be seen that the completeness is
larger than 80\%
for $W_\mathrm{r}(\lambda 1560)\ge 0.4$~\AA.
%The typical S/N ratio of the spectra is decreasing towards
%longer wavelengths such that the above completeness limit is actually
%only valid for spectra with {\it median} S/N ratios per pixel of seven.

\subsection{Number of absorbers per unit redshift}

\label{sec:nz}

\begin{figure}[htb]
\resizebox{\hsize}{!}{\includegraphics[angle=90.]{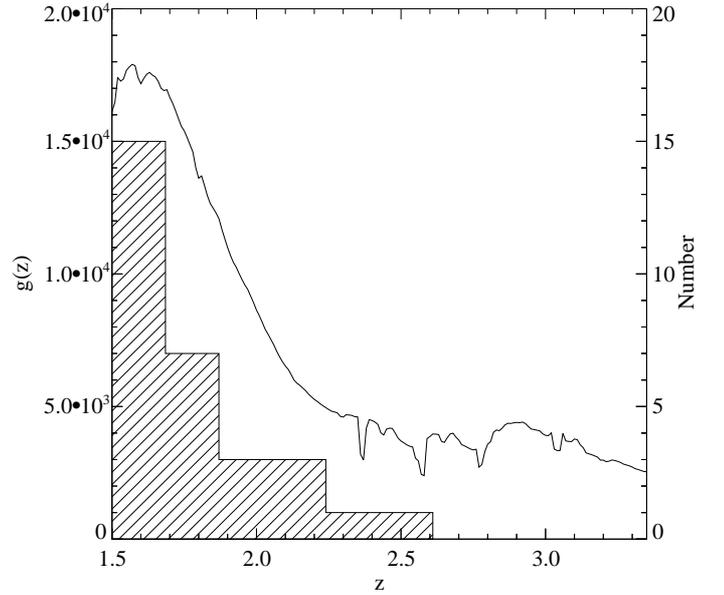}}
\caption{Statistical sensitivity function of our \ion{C}{i} survey, $g(z)$ (solid curve; left-hand side
axis). The dips seen at  $z\sim 2.35$ (resp. 2.6) are due to \ion{C}{i}\,$\lambda$1656
(resp. \ion{C}{i}\,$\lambda$1560) falling over bad pixels at the junction between the
two SDSS CCD chips. Dips at higher redshifts ($z\sim 2.8,3.05$) are due to telluric
lines (e.g., [\ion{O}{i}]\,$\lambda$6300). The hashed histogram (right-hand side axis)
shows the \ion{C}{i}-absorber number counts in different redshift bins for intervening
systems with rest-frame equivalent width above the completeness limit of the
survey, i.e., $W_\mathrm{r}(\lambda 1560)\ge 0.4$~\AA.\label{fig:newgz}}
\end{figure}

We calculated the sensitivity function, $g(z)$, of our survey, i.e., the number of
lines-of-sight probing a given redshift $z$ and having $\mathrm{S/N}>4$ at the expected positions of
both \ion{C}{i} lines. This function is shown in Fig.~\ref{fig:newgz}. It
combines together the [$z_\mathrm{min}$,$z_\mathrm{max}$] pairs, previously defined
in Sect.~\ref{sec:identification}, for all the lines-of-sight. We further excluded
the regions with velocities relative to the QSO emission redshifts smaller than
$5000$~km\,s$^{-1}$, which could, in principle, be influenced by the quasar (see
Sect.~\ref{sec:prox}). Note that the uncertainties on $z_\mathrm{em}$ are of the order of
$500$~km\,s$^{-1}$ \citep[see, e.g.,][]{2012A&A...548A..66P} and therefore are small
enough not to affect the statistics. The total statistical absorption path length probed by the
QSO sample over $z=1.50-3.35$ is $\Delta z\approx 13\,000$ with an average redshift
$\avg{z}=1.9$. From Fig.~\ref{fig:newgz}, it is apparent that the sensitivity of the
survey is an increasing function towards lower redshifts. One therefore expects a
larger number of absorbers to be found at $z<2$. This is what is observed in practice as
indicated by the redshift histogram of the detected intervening \ion{C}{i} absorbers
over-plotted on the same figure. On the other hand, the small number of \ion{C}{i}
systems found at $z_\mathrm{abs}>2.2$ (i.e., eight systems out of
a total of 66 systems or, equivalently, 12\% of the sample) is striking.

\begin{figure}[htb]
%\centering{\hbox{
%\psfig{figure=dNdz.ps,width=8.9cm,clip=,bbllx=25.pt,bblly=115.pt,bburx=543.pt,bbury=704.pt,angle=90.}}}
\resizebox{\hsize}{!}{\includegraphics[angle=90.]{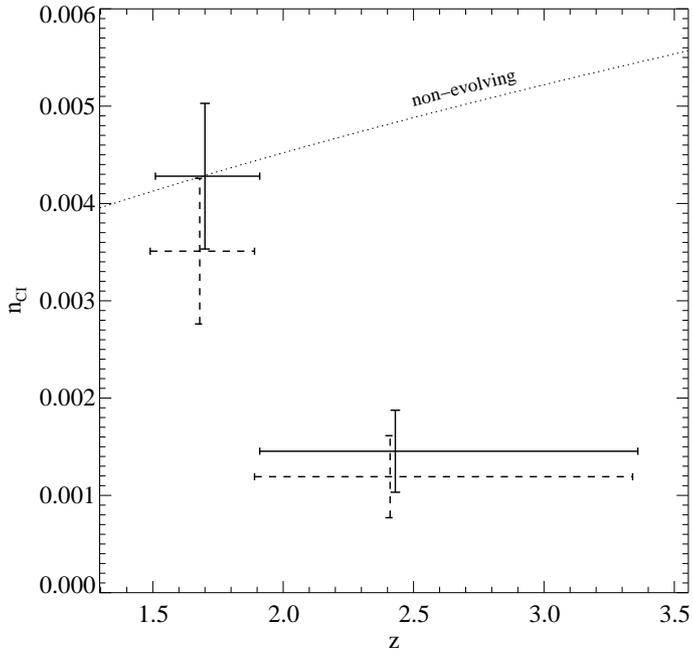}}
\caption{Evolution of the number of intervening \ion{C}{i} absorbers per
unit redshift, $n_\mathrm{\ion{C}{i}}$, for systems
with rest-frame equivalent widths $W_\mathrm{r}(\lambda 1560)\ge 0.4$~\AA.
{\bf The measurements (solid) are corrected for incompleteness at the
latter limit. For clarity purposes, the uncorrected data (dashed) are displayed
with a slight offset in abscissa.} The dotted curve shows the expected redshift
behaviour of a non-evolving population.\label{fig:dndz}}
\end{figure}

To investigate this further, we calculated the number of intervening \ion{C}{i} absorbers
per unit redshift, $n_\mathrm{\ion{C}{i}}$, in two redshift bins of roughly equal total
absorption path length (with boundary redshift $z = 1.9$). Here, we only considered the systems with
rest-frame equivalent widths above the completeness limit of the
survey, i.e., $W_\mathrm{r}(\lambda 1560)\ge 0.4$~\AA. The results are
summarised in Table~\ref{tab:dNdz} and shown
in Fig.~\ref{fig:dndz}. In the table, we also separated
the strongest from the weaker absorbers (around a median rest-frame equivalent width of
0.64~\AA) but there is no obvious difference between the redshift evolution of these two groups.

% Using all spectra with median $\mathrm{S/N}>4$, the relative fraction of
% \ion{C}{i} systems in the higher redshift bin is smaller than when
% using spectra of higher S/N ratios. We checked this by separating
% the strongest from the weaker systems (the median value is
% $W_\mathrm{r}(\lambda 1560)=0.64$~\AA). This shows that the weaker
% systems are being missed at higher redshifts. This can easily be
% understood as the S/N ratio tends to be lower at longer wavelengths
% while we are selecting the spectra based on their median S/N
% ratio. This implies that robust results on a possible redshift
% evolution of the \ion{C}{i} systems can either be obtained with
% $\mathrm{S/N}>4$ spectra considering only the strongest absorbers (having
% $W_\mathrm{r}(\lambda 1560)\ge 0.64$~\AA) or, equivalently, with
% $\mathrm{S/N}>7$ spectra considering stronger and weaker systems
% altogether. Given the larger number statistics, we adopt the latter
% results and show them in Fig.~\ref{fig:dndz} (right panel).
% % This possible evolution is real as the median S/N ratio per pixel
% % is similar in both redshift sub-samples (see the last column of
% % Table~\ref{tab:dNdz}).

The $n_\mathrm{\ion{C}{i}}\sim 1.4\times 10^{-3}$ we measure in the higher redshift
bin ($1.9<z_\mathrm{abs}<3.35$),
taking into account the effect of incompleteness estimated in Sect.~\ref{sec:comp},
implies that \ion{C}{i} systems with $W_\mathrm{r}(\lambda 1560)\ge 0.4$~\AA\
are more than one hundred-times
rarer than DLAs at $z_\mathrm{abs}=2.5$ \citep[see][]{2012A&A...547L...1N}.
An evolution of $n_\mathrm{\ion{C}{i}}$, with nearly thrice as many systems
below $z=1.9$ than above that, is also observed. Compared to the redshift behaviour of
a non-evolving population, this is significant at the
$4.3\sigma$ level. Such an evolution is interesting and should be studied further as
it depends on the balance between dust shielding and the UV radiation field. This may
imply a strong evolution of the shielding of 10~eV photons by dust between $z=2.5$
and $z=1.5$.

\subsection{Proximate systems}

\label{sec:prox}

% Hence, 21\% of the systems in the overall sample could be associated
% with the QSO.

There are 14 \ion{C}{i} systems with velocities relative to the QSO
emission redshifts smaller than $5000$~km\,s$^{-1}$. These could
be associated with the QSO host galaxy or nearby
environment. Six systems even have absorption redshifts larger than
the corresponding QSO emission redshifts (by up to $\sim
4000$~km\,s$^{-1}$) which is difficult to explain by large peculiar
velocities in intervening systems. Imposing the same data-quality cuts
and minimum equivalent widths as in the previous section, we find that
the incidence of \ion{C}{i} absorbers at small velocity differences
from the quasars is consistent with that of intervening systems.
However, associated errors are large due to small number statistics.

Because of the clustering of galaxies around the massive QSO host
galaxies, an excess of proximate \ion{C}{i} systems could be
expected. However, \ion{C}{i} is a fragile species which can easily be
photo-ionised by the intense UV radiation emitted by the
QSO engine. Interestingly, the lack of a significant excess of proximate
systems was also observed by \citet{2008ApJ...675.1002P} considering DLAs. In
this case, the abundance of proximate DLAs is only a factor of two
larger than that of the overall DLA population. A similarly low
over-abundance factor was observed by \citet{2013A&A...558A.111F} for
strong DLAs with $\log N(\ion{H}{i})>21.3$ (atoms~cm$^{-2}$). This is
much less than what is
expected based on clustering arguments alone.

In this work, we do not observe that the properties of proximate
\ion{C}{i} systems are different from those of intervening \ion{C}{i} systems. This
is true for redshift, \ion{C}{i} equivalent-width, \ion{H}{i} content,
reddening and UV bump-strength distributions. Nevertheless, because of
the possibly different origin of these absorbers and the possible requirement of
strong dust shielding from the nearby QSOs, we discriminate in the following
proximate \ion{C}{i} systems from the rest of the
population and comment, whenever possible, on proximate \ion{C}{i} systems of interest.

\section{Evidence for dust}

\subsection{QSO optical colours}

\label{sec:colours}

In order to assess the impact of the \ion{C}{i} absorbers on their
background QSOs and check for the existence of dust in these systems,
we first consider the observed colours of these QSOs and compare them
with the colours of the overall QSO population used as a control
sample.

\begin{figure*}[htb]
\centering{\hbox{
\psfig{figure=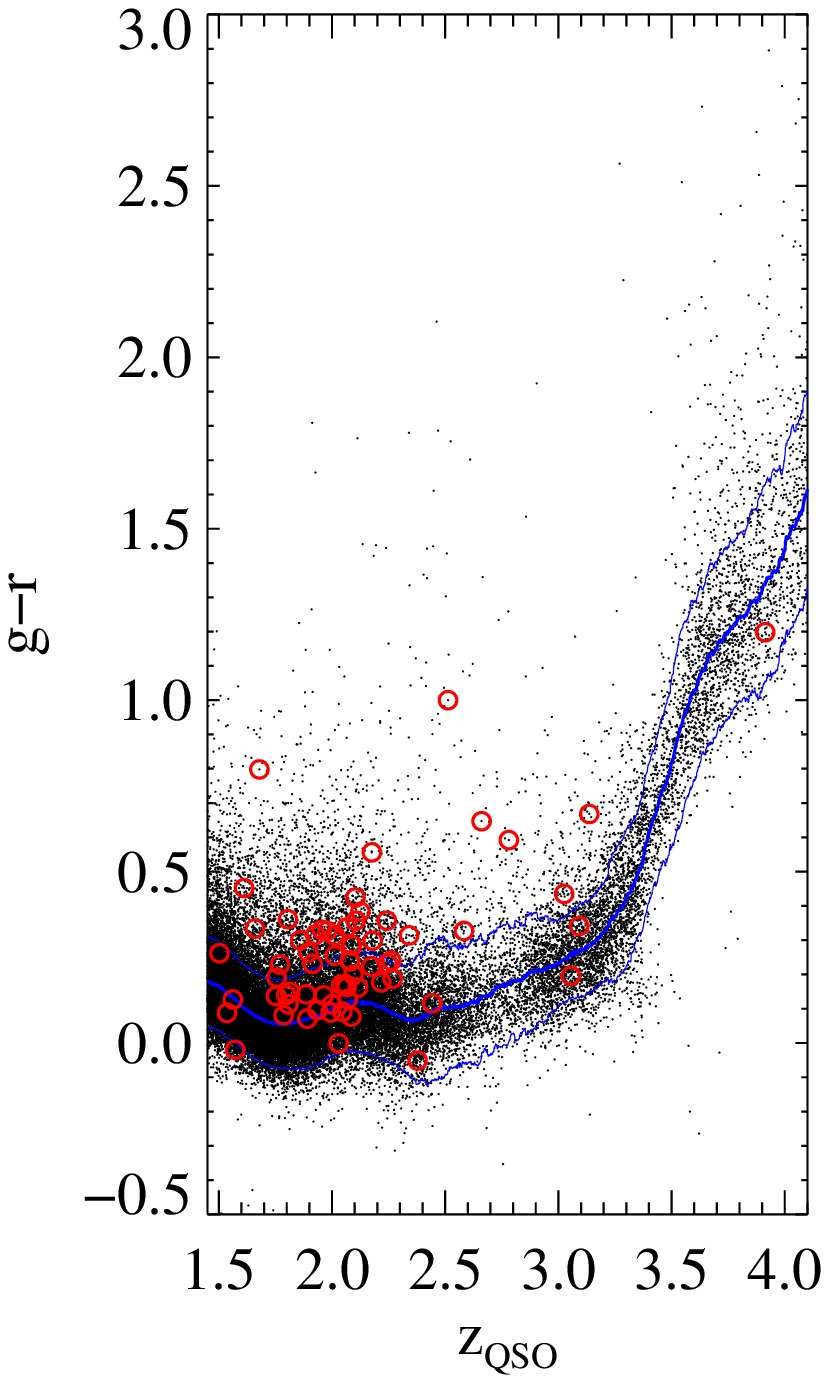,width=5.8cm,clip=,bbllx=70.pt,bblly=176.pt,bburx=318.pt,bbury=573.pt,angle=0.}\hspace{+0.3cm}
\psfig{figure=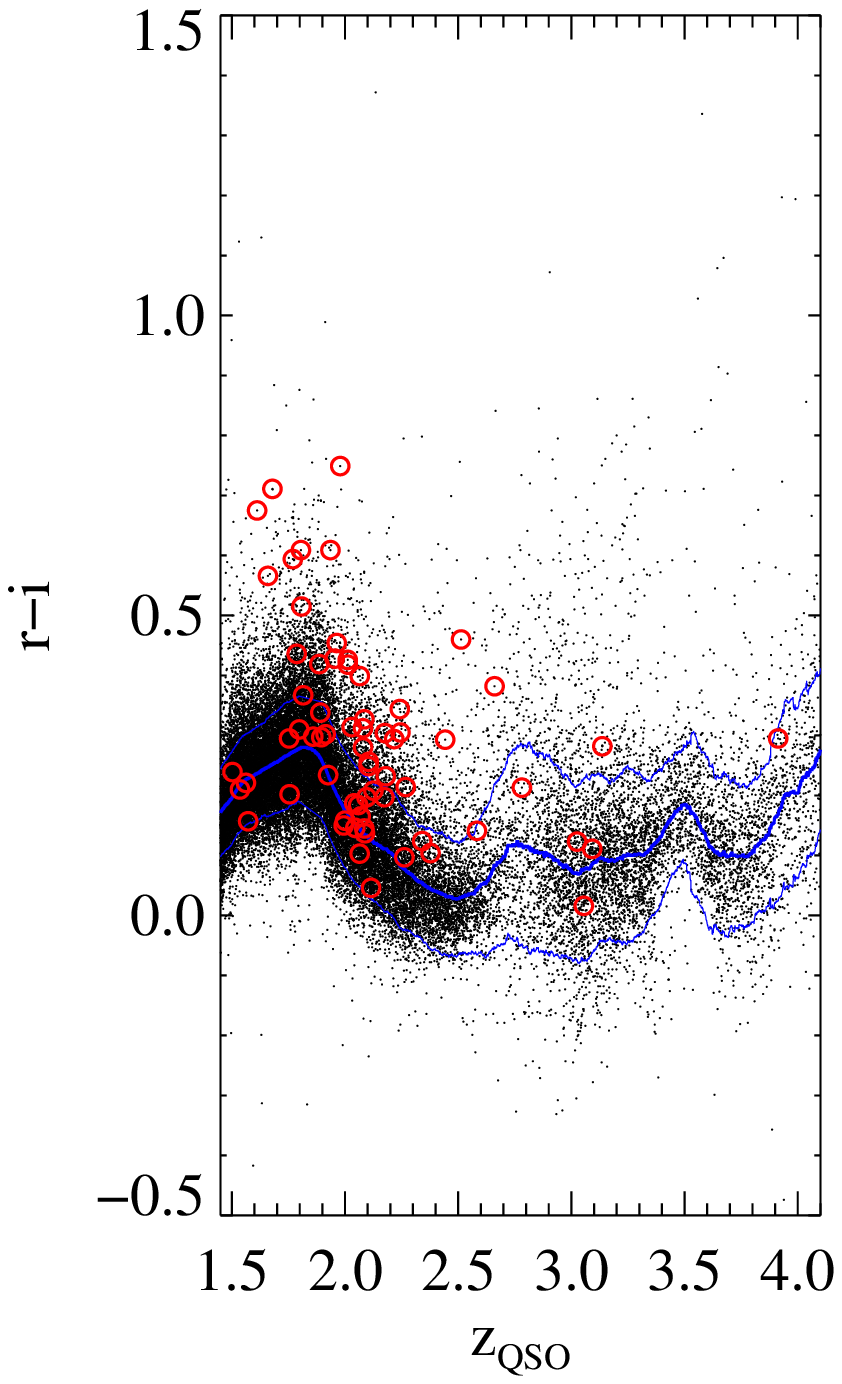,width=5.8cm,clip=,bbllx=70.pt,bblly=176.pt,bburx=318.pt,bbury=573.pt,angle=0.}\hspace{+0.3cm}
\psfig{figure=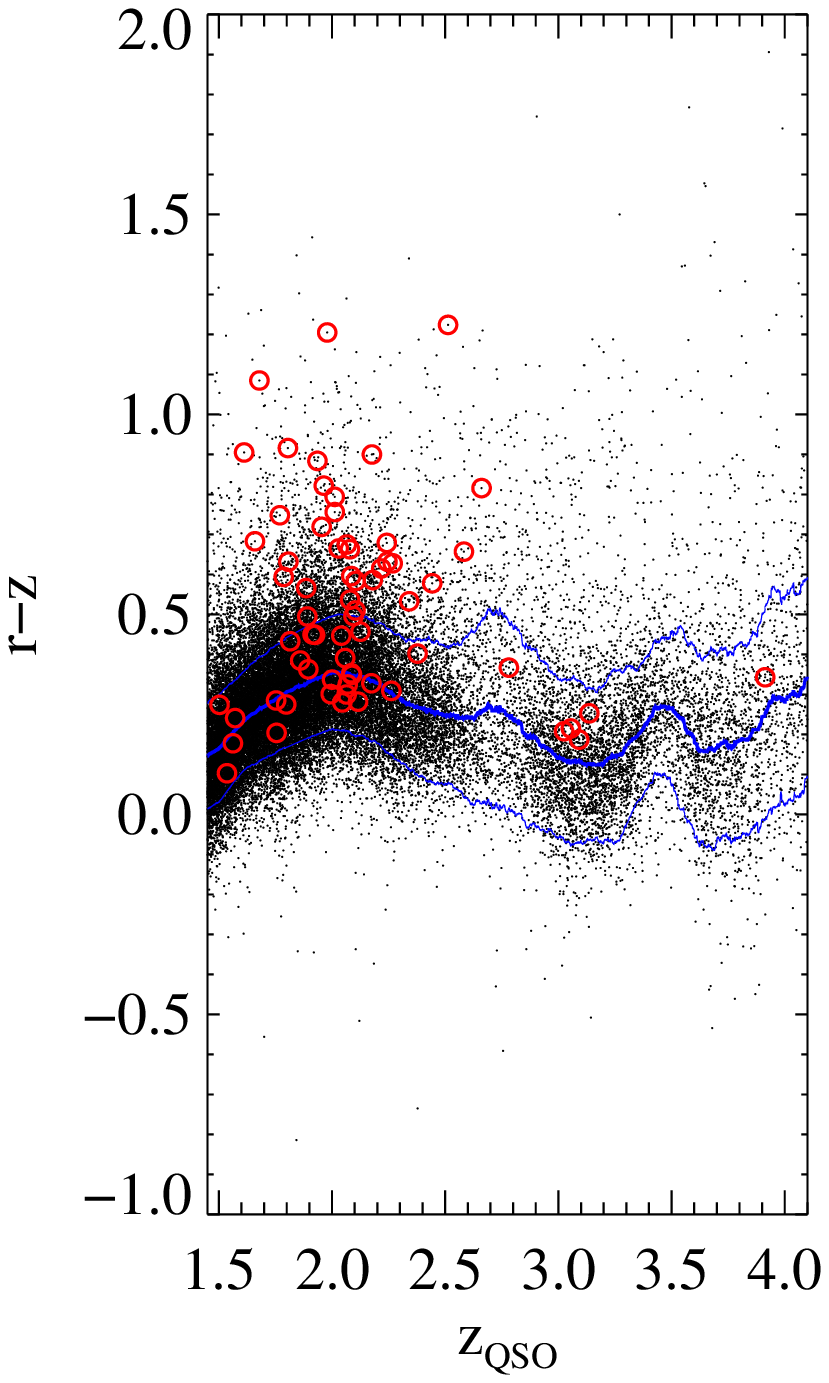,width=5.8cm,clip=,bbllx=70.pt,bblly=176.pt,bburx=318.pt,bbury=573.pt,angle=0.}}}\vspace{+0.3cm}
\centering{\hbox{
\psfig{figure=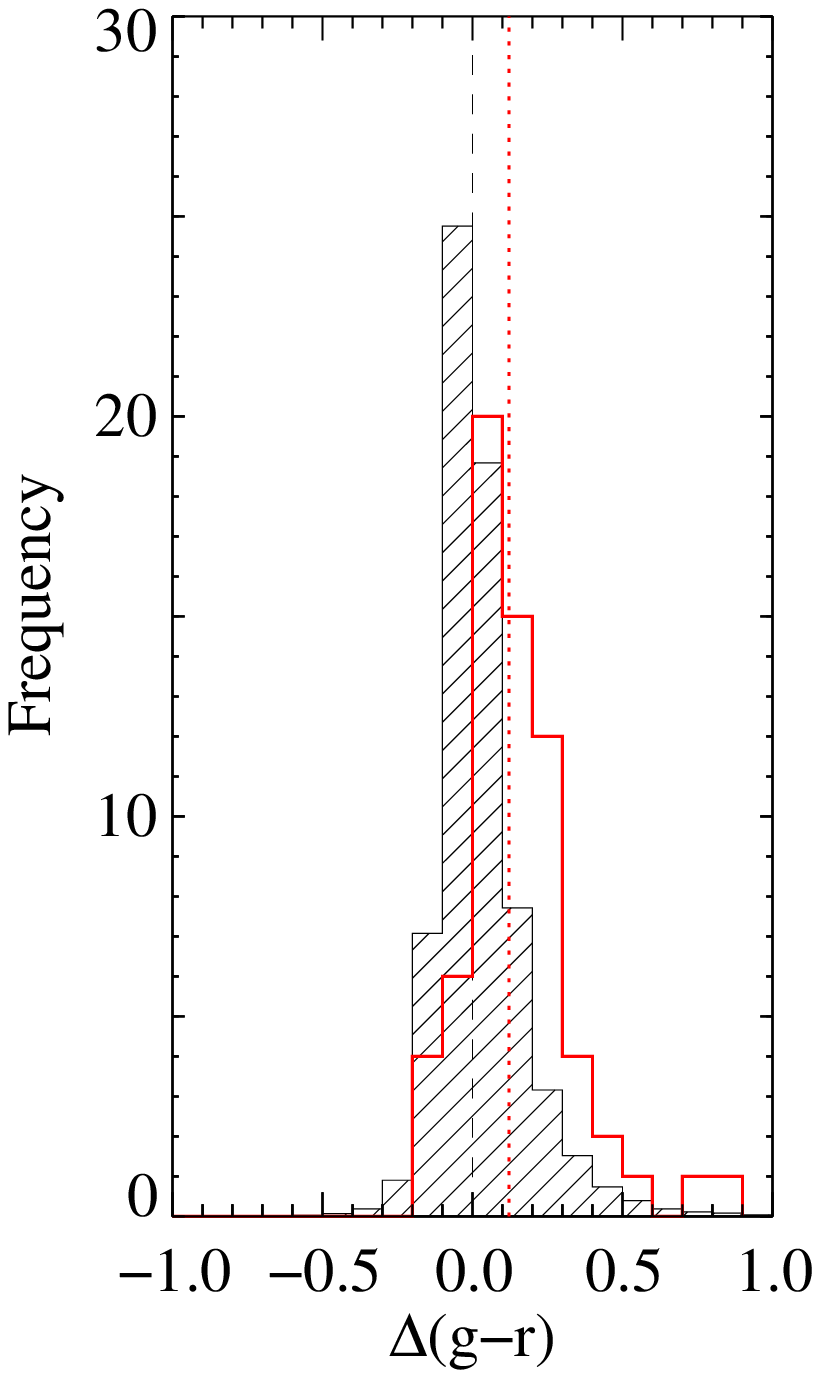,width=5.8cm,clip=,bbllx=70.pt,bblly=176.pt,bburx=318.pt,bbury=573.pt,angle=0.}\hspace{+0.3cm}
\psfig{figure=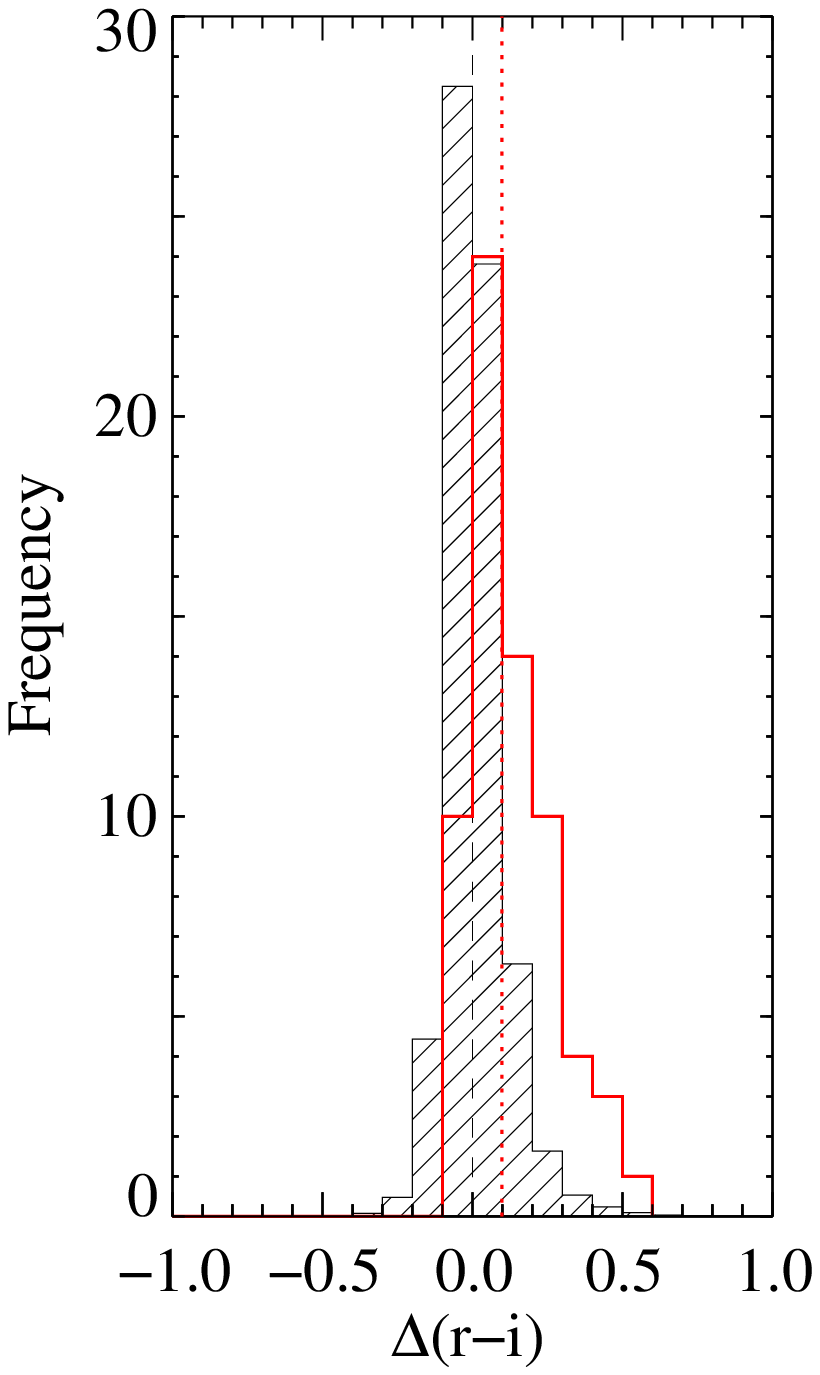,width=5.8cm,clip=,bbllx=70.pt,bblly=176.pt,bburx=318.pt,bbury=573.pt,angle=0.}\hspace{+0.3cm}
\psfig{figure=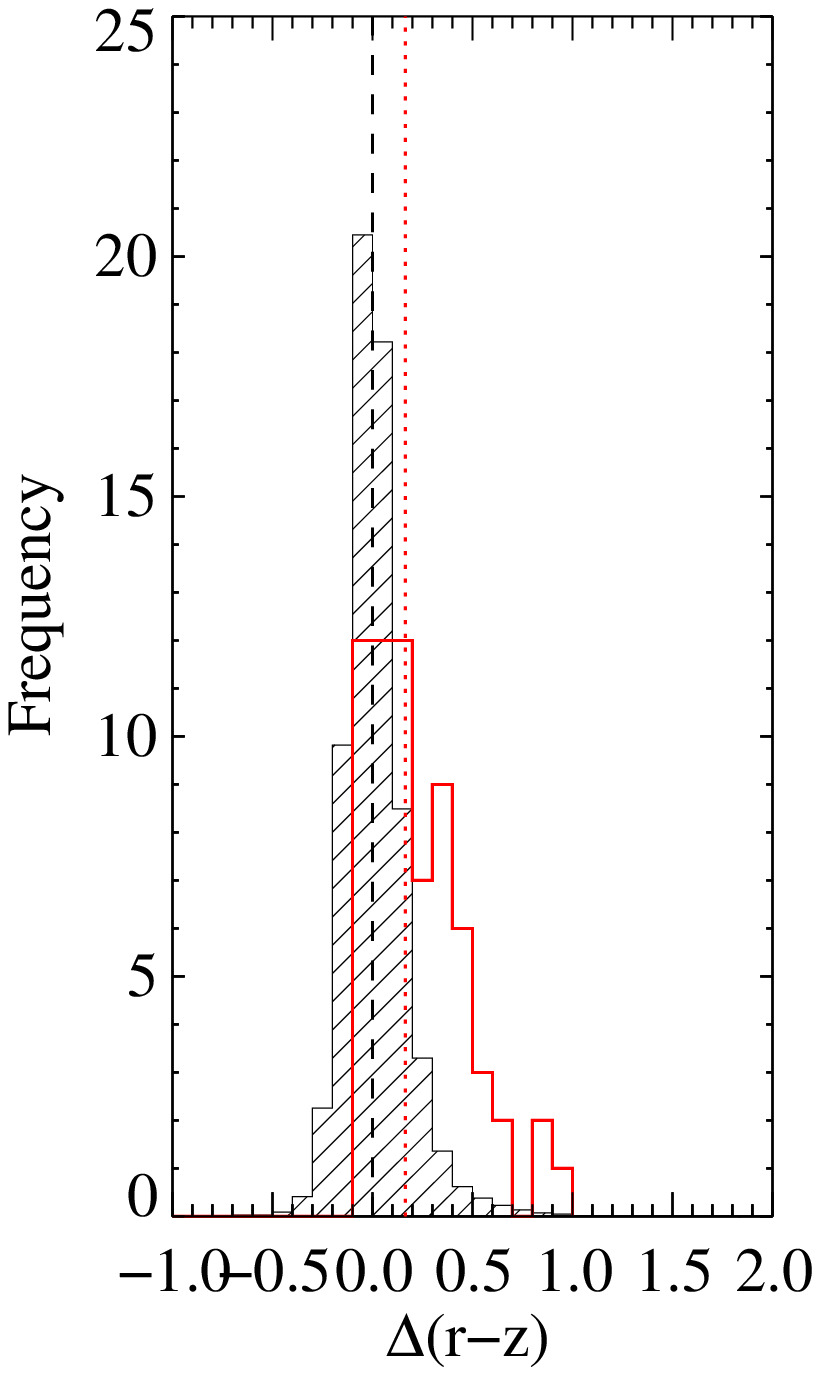,width=5.8cm,clip=,bbllx=70.pt,bblly=176.pt,bburx=318.pt,bbury=573.pt,angle=0.}}}
\caption{{\it Upper panels:} Colours of the 41\,696 SDSS QSOs whose
  spectra were searched for \ion{C}{i} absorption versus QSO emission
  redshift. $(g-r)$, $(r-i)$ and $(r-z)$ colours are displayed in the
  left, middle and right-hand side panels, respectively. The thick
  blue line shows the median QSO colour as a function of redshift. Thin
  blue lines are shown one standard deviation away from that median. The
  red circles indicate the 66 lines-of-sight passing through
  \ion{C}{i}-bearing gas. {\it Lower panels:} Distribution of colour
  excesses [defined as, e.g.: $(g-r)-\avg{g-r}_{z_\mathrm{QSO}}$ for
  $\Delta(g-r)$] for the 66 QSOs whose spectra were found to exhibit
  \ion{C}{i} absorption (thick red histogram) compared to the whole
  QSO sample (thin-hashed histogram, scaled down by a factor of 632
  to have the same area). In each panel [from left to right: $\Delta(g-r)$,
  $\Delta(r-i)$ and $\Delta(r-z)$], vertical dotted red lines indicate
  the median colour excesses of the sample of QSOs with detected
  \ion{C}{i} absorption. The median values for the
  whole QSO sample are shown by vertical dashed lines.\label{fig:colours}}
\end{figure*}

In Fig.~\ref{fig:colours}, we show the distributions of $(g-r)$,
$(r-i)$ and $(r-z)$ colours for the 41\,696 QSOs whose spectra were
searched for \ion{C}{i} absorption. In the upper panels of this
figure, it is apparent that the lines-of-sight with detected
\ion{C}{i} absorption do not distribute in the same way as the other
lines-of-sight. They are displaced altogether towards redder
optical colours compared to the average loci of the QSO redshift
sequences. The effect is most easily seen in the lower panels of
Fig.~\ref{fig:colours}, which compare the colour histograms of the
two QSO populations (i.e., the \ion{C}{i}-detected QSO sample and the
overall QSO sample). The two-sided Kolmogorov-Smirnov test probability
that the two samples are drawn from the same parent distribution is as
small as $\la 10^{-10}$. The typical colour excess is $\sim 0.15$~mag,
i.e., about five times larger than the mean $(r-z)$ colour excess of
0.03~mag derived by \citet{2008A&A...478..701V} in
$z_\mathrm{abs}\approx 2.8$ DLAs from SDSS DR\,5. A similar result for
DLAs was found by \citet{2012MNRAS.419.1028K} based on $(g-i)$ colours
of SDSS-DR\,7 QSOs. This is clear evidence for the presence of dust
among \ion{C}{i} absorbers.

\subsection{Reddening}

\label{sec:ebv}

Motivated by the unequivocal signature of dust in the form of a colour
excess of the background QSOs with detected \ion{C}{i} absorption, we
now aim at constraining the properties and the nature of dust in these
systems.

\begin{figure*}[htb]
%\centering{\hbox{
%\psfig{figure=CI_sed_fitting.ps,width=18.35cm,clip=,bbllx=120.pt,bblly=22.pt,bburx=484.pt,bbury=745.pt,angle=90.}}}
\resizebox{\hsize}{!}{\includegraphics[angle=90.]{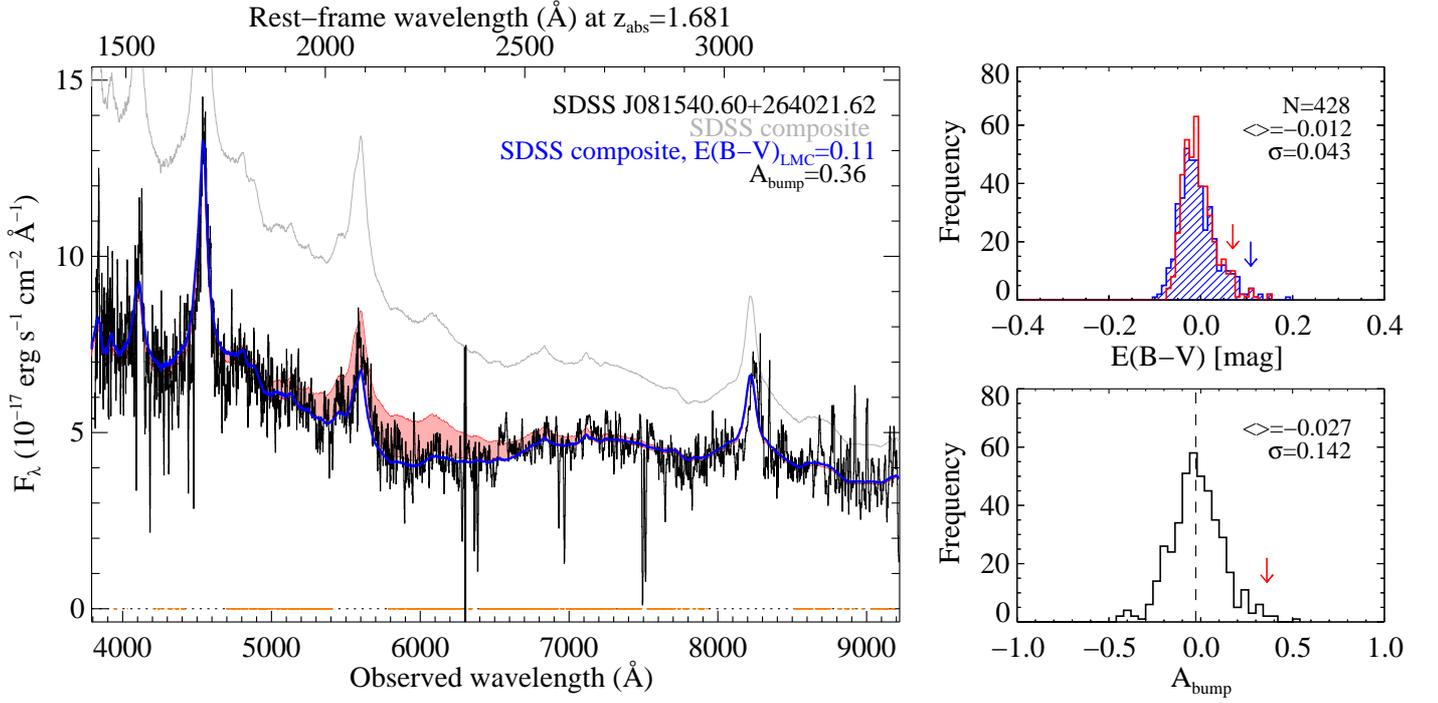}}
\caption{{\it Left panel:} Illustration of reddening measurement based on
  QSO SED fitting. The blue curve shows our best fit to QSO continuum using
  the QSO template spectrum from \citet{2001AJ....122..549V} and the best-fitting
  LMC extinction law at the redshift of the \ion{C}{i} absorber. The
  non-extinguished QSO template is displayed in grey. The additional presence of a
  2175~\AA\ absorption feature is shown by the light-red shaded area
  (i.e., $A_\mathrm{bump}$). The broken orange line at zero ordinate
  indicates the wavelength regions devoid of QSO emission lines, strong
  absorption lines and sky residuals, which were used during the $\chi^2$
  minimization. {\it Upper right panel:} Distribution of reddening
  measured in the QSO control sample assuming either a SMC (red) or
  the best-fitting extinction law (in this case, LMC; blue). The
  reddening values of the sight-line under consideration
  are indicated by downward arrows. {\it Lower right panel:} Same as
  above but for the distribution of 2175~\AA\ bump
  strengths.\label{fig:sed}}
\end{figure*}

For each of the 66 QSOs with foreground \ion{C}{i} absorbers, we
derived the QSO reddening, $E($B-V$)$, following the same approach
as used in, e.g., \citet{2008MNRAS.391L..69S} and
\citet{2009A&A...503..765N,2010A&A...523A..80N}. First, we corrected
the QSO spectra for Galactic reddening using the extinction maps from
\citet{1998ApJ...500..525S}. We then fitted the spectra with the SDSS
QSO composite spectrum from \citet{2001AJ....122..549V} shifted to
that QSO emission redshift and reddened with either a Small Magellanic
Cloud (SMC), Large Magellanic Cloud (LMC), LMC2 super-shell or Milky
Way (MW) extinction law \citep{2003ApJ...594..279G} at the
\ion{C}{i}-absorber redshift. Our procedure is illustrated in the left
panel of Fig.~\ref{fig:sed}. The fit with the smallest $\chi^2$ value
indicates the most representative extinction law for a given
absorber. The latter is specified in Col. 'Best fit' of
Table~\ref{tab:sed} and the corresponding $E($B-V$)$ value is given
in the preceding column.

For each QSO line-of-sight exhibiting \ion{C}{i} absorption, we defined
a control sample (hereafter denoted as ``C.S.'') made of SDSS-DR\,7 QSOs
from the searched sample having an emission redshift within $\pm 0.05$
and a $z$-band magnitude within $\pm 0.1$~mag from those of the QSO under
consideration. In some instances, this resulted in a sample of less than 30 QSOs
in which case we increased the above maximum magnitude difference by steps
of 0.01 until the number of QSOs in the control sample reached (or exceeded)
30. We then applied to each QSO spectra from the control sample the exact
same fitting procedure as described in the previous
paragraph. Table~\ref{tab:sed} lists the number of QSOs and the median reddening
and standard deviation of the distribution of $E($B-V$)$ values in each
control sample (see the upper right panel of Fig.~\ref{fig:sed} for an
illustration). The values given in Table~\ref{tab:sed} correspond to the
most representative extinction law previously determined for that
particular \ion{C}{i}-detected QSO spectrum.

\begin{figure*}[htb]
%\centering{\hbox{
%\psfig{figure=histebv_CI.ps,width=9.1cm,clip=,bbllx=82.pt,bblly=174.pt,bburx=491.pt,bbury=573.pt,angle=0.}
%\psfig{figure=hist_bump_CI.ps,width=9.1cm,clip=,bbllx=82.pt,bblly=174.pt,bburx=491.pt,bbury=573.pt,angle=0.}}}
\resizebox{\hsize}{!}{\includegraphics{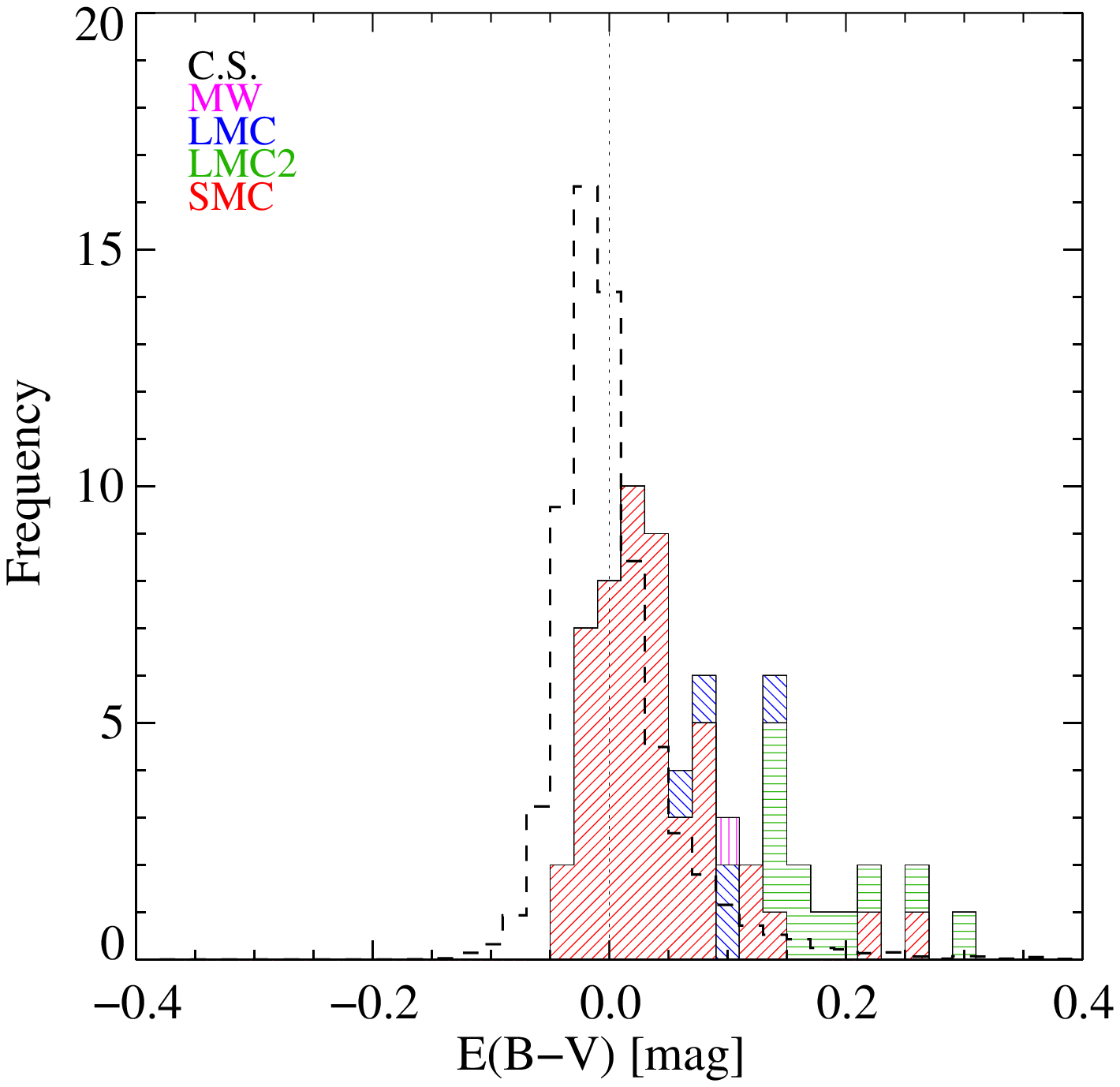}\includegraphics{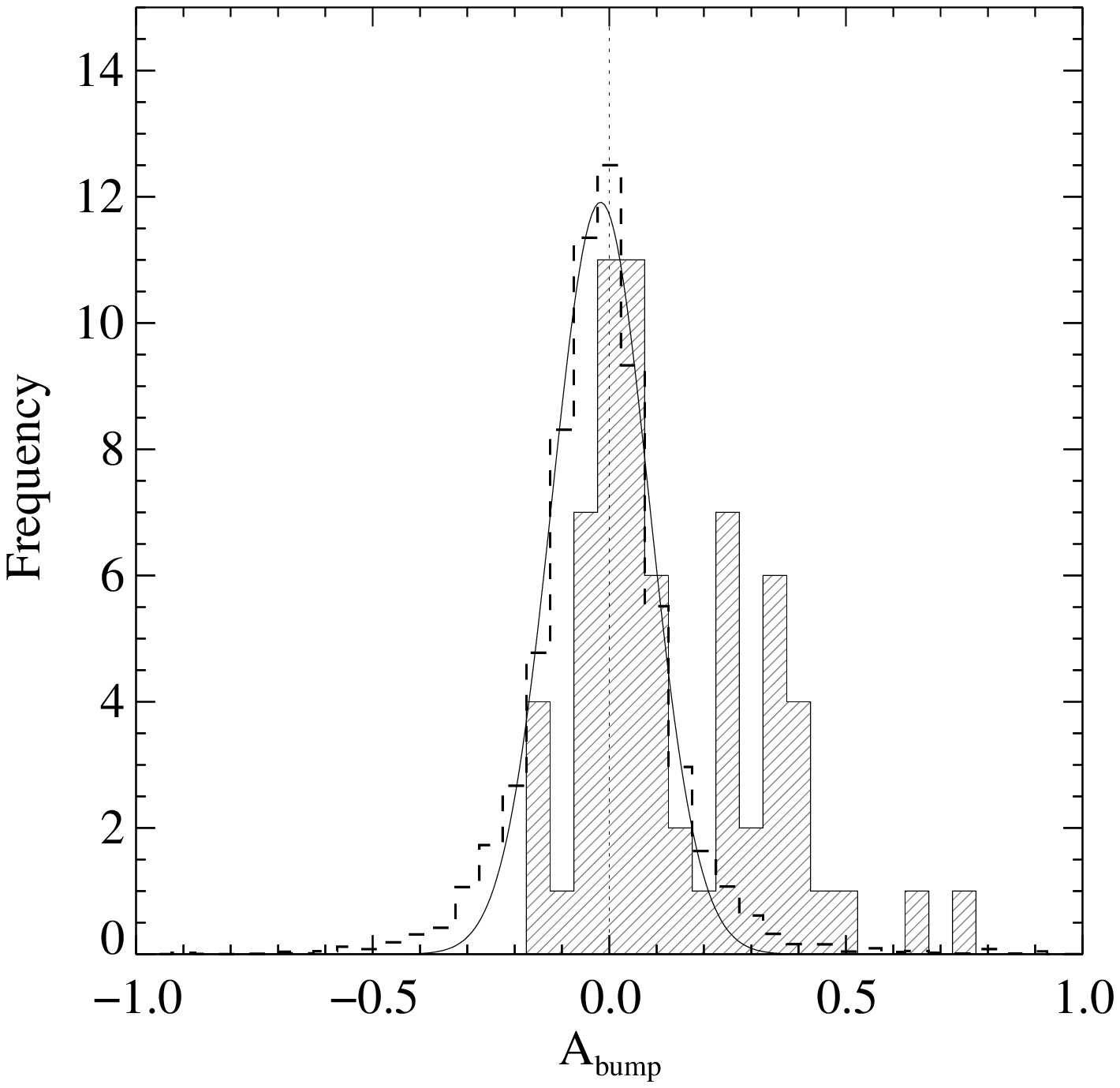}}
\caption{{\it Left panel:} Histogram of reddening for the sample of
  66 QSOs whose spectra were found to exhibit \ion{C}{i} absorption
  (multi-color). The colour/hash coding for the different
  \ion{C}{i}-detected lines-of-sight relates to the best-fitting
  extinction law (red: SMC; blue: LMC; green: LMC2; purple: MW). The
  distribution of reddening from the QSO control sample (C.S.), i.e., the
  sum of the normalized distributions of individual QSO control samples,
  is displayed with dashed lines. {\it Right panel:} Same as in the left
  panel but for the histogram of 2175~\AA\ bump strengths. The
  distribution from the QSO control sample is well represented by a
  Gaussian profile (solid curve).\label{fig:histo}}
\end{figure*}

In the left panel of Fig.~\ref{fig:histo}, we show the histogram of
reddening for the sample of 66 QSO lines-of-sight with detected
\ion{C}{i} absorbers compared to the cumulative control sample
(calculated as the sum of the normalized distributions of individual
control samples). As in Sect.~\ref{sec:colours}, an offset between the
two samples is apparent. The mean reddening induced by \ion{C}{i}
systems is 0.065~mag. A tail in the histogram of the
\ion{C}{i}-detected lines-of-sight is observed, with $E($B-V$)$
values up to $\sim 0.3$~mag.

\subsection{The 2175~\AA\ extinction feature}

\label{sec:abump}

A number of \ion{C}{i}-detected QSO spectra are best-matched by an
extinction law exhibiting the absorption feature at rest-frame wavelength 2175~\AA. In order
to measure the strength of this UV bump (denoted $A_\mathrm{bump}$),
we followed a prescription similar to the one used by
\citet{2010ApJ...720..328J} where the observed QSO spectrum was fitted
with the SDSS QSO composite spectrum reddened via a parametrized
pseudo-extinction law made of a smooth component and a Drude
component. However, here we fixed the wavelength and width of the bump
to the Galactic values determined by \citet{2007ApJ...663..320F}. Both
of these quantities indeed show little variation from line-of-sight to
line-of-sight through the Galaxy and the Magellanic Clouds. This then
limits the number of free parameters and prevents the fit from
diverging towards very wide and shallow solutions which could be
non-physical. Indeed, imperfect matching of the observed QSO continuum
by the smooth component is expected due to intrinsic QSO-shape
variations \citep[e.g.,][]{2000PASP..112..537P}.

The fitting process is illustrated in the left panel of
Fig.~\ref{fig:sed}. The shaded area represents the measure of the bump
strength. This is the difference between the above best-fit function
and the same function but considering only its smooth component (i.e.,
with the Drude component set to zero). $A_\mathrm{bump}$ values are
listed for each absorber in Table~\ref{tab:sed}. As previously done
for the determination of reddening (see Sect.~\ref{sec:ebv}), we also
defined a QSO control sample whose measured $A_\mathrm{bump}$
distribution is shown in the lower right panel of Fig.~\ref{fig:sed}
(i.e., for the given QSO emission redshift). Table~\ref{tab:sed} gives
for each control sample the median and standard deviation of this
distribution.

The histogram of bump strengths in the \ion{C}{i}-absorber sample is
displayed in the right panel of Fig.~\ref{fig:histo}. One can see from
this figure that more than a quarter of the \ion{C}{i} systems feature
absorption at 2175~\AA. This strengthens the result from the previous
section that significant reddening of the background QSOs by dust is induced
by some of the \ion{C}{i} absorbers. We will come back to this and quantify the
effect in Sect.~\ref{sec:relations}.

In the following, we shall use the median $E($B-V$)$ and $A_\mathrm{bump}$
values of the control samples, i.e., $\langle E($B-V$)\rangle_\mathrm{C.S.}$
and $\langle A_{\rm bump}\rangle_\mathrm{C.S.}$, to define the exact colour excess and
bump strength towards a given \ion{C}{i}-detected QSO line-of-sight:
$E($B-V$)=E($B-V$)_\mathrm{measured}-\langle E($B-V$)\rangle_\mathrm{C.S.}$,
and likewise for $A_\mathrm{bump}$. These zero-point corrections are usually almost
negligible (see Table~\ref{tab:sed}). In addition, the standard deviations of
$E($B-V$)$ and $A_\mathrm{bump}$ values in each control sample provide
an estimate of the uncertainty due to intrinsic QSO-shape
variations \citep[see][]{2000PASP..112..537P} and hence the significance of the
reddening induced by each \ion{C}{i} absorber and the significance of associated
2175~\AA\ absorption, respectively.

\section{\ion{H}{i} content}

\label{sec:nhi}

\begin{figure}[htb]
%\centering{\hbox{
%\psfig{figure=histHI.ps,width=8.9cm,clip=,bbllx=25.pt,bblly=115.pt,bburx=543.pt,bbury=704.pt,angle=90.}}}
\resizebox{\hsize}{!}{\includegraphics[angle=90.]{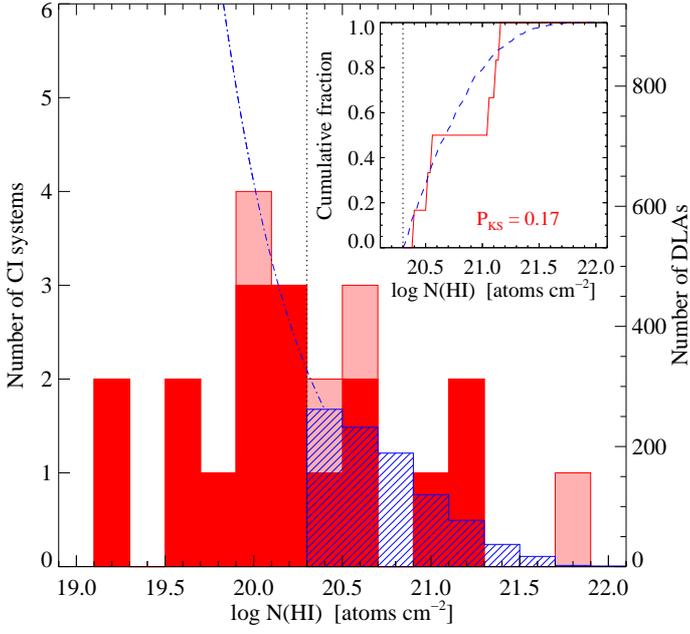}}
\caption{Neutral atomic-hydrogen column-density distribution of
  \ion{C}{i} absorbers (red-filled histogram; left-hand side
  axis). Proximate systems are displayed in salmon. The $N(\ion{H}{i})$
  distribution of intervening DLAs from SDSS DR7 \citep[][blue-hashed
  histogram]{2009A&A...505.1087N} is over-plotted using a different
  scaling (right-hand side axis) so that the areas of the two
  histograms above $\log N(\ion{H}{i})=20.3$ are equal. The blue
  dashed-dotted curve is a fit of the distribution in the sub-DLA
  regime from \citet{2014MNRAS.438..476P}. The inset shows the
  cumulative $N(\ion{H}{i})$ distributions of the intervening \ion{C}{i}
  and DLA samples starting at $\log N(\ion{H}{i})=20.3$ (atoms~cm$^{-2}$)
  and the two-sided Kolmogorov-Smirnov test probability that the two
  distributions come from the same parent population.\label{fig:histHI}}
\end{figure}

As part of a spectroscopic campaign which we will describe in a
companion paper, we followed up the \ion{C}{i} absorbers from the
overall sample which are observable from the southern hemisphere using
VLT/UVES. We present in Fig.~\ref{fig:histHI}
the \ion{H}{i} column-density distribution of this
\ion{C}{i}-absorber sub-sample (referred to in the following as the
\ion{H}{i} sub-sample) and compare it with the distribution of $N(\ion{H}{i})$ from
systematic DLA and/or sub-DLA surveys.

We secured \ion{H}{i} column-density measurements for most of the
systems in the overall sample which have a declination of
$\delta <+28\deg$, i.e., 14 out of 16 systems at redshifts
$z_\mathrm{abs}>1.8$ (the two exceptions being the lines-of-sight
towards SDSS~J\,091721.37$+$015448.1 and J\,233633.81$-$105841.5)
and four out of eight systems at $z_\mathrm{abs}\approx 1.75$ (see
Table~\ref{tab:ci}). While \ion{H}{i} column densities derived from UVES
spectroscopy are usually more accurate, the last two columns of
Table~\ref{tab:ci} show that they confirm those derived
directly from SDSS spectra as testified by the five systems at
$z_\mathrm{abs}>2.2$ where this measurement could be done from both
datasets. For this reason, we here complement our UVES measurements
with the values we derived using SDSS spectra for the three systems at
$z_\mathrm{abs}>2.2$ for which high-resolution spectroscopic data are
not available because the background QSOs are too far North for the
VLT to observe them. The \ion{H}{i} sub-sample thus comprises a total of 21
systems.

In Fig.~\ref{fig:histHI}, we compare the observed \ion{H}{i} column-density
distribution of \ion{C}{i}-selected absorbers with that of
\ion{H}{i}-selected DLAs (from SDSS DR\,7 as well;
\citealt{2009A&A...505.1087N}). In this figure, we also show the
expected number of sub-DLAs using the fitted distribution function
from \citet{2014MNRAS.438..476P}. We find that a large fraction of the
\ion{C}{i} absorbers have neutral atomic-hydrogen column densities
slightly below the conventional DLA limit ($N(\ion{H}{i})= 2\times 10^{20}$
atoms cm$^{-2}$) and therefore classify as strong sub-DLAs. However,
the fraction of \ion{C}{i} absorbers among sub-DLAs is much less than
among DLAs indicating that efficient shielding is much more difficult
to obtain below the DLA limit. Though rare, the existence of
\ion{C}{i} absorbers with low neutral atomic-hydrogen column densities
supports the presence of dust in these systems. The dust-to-gas ratio
in these systems has to be high enough so that the absorption of UV
photons by dust allows \ion{C}{i} to be present in large amounts.

No \ion{C}{i} system is found with $\log N(\ion{H}{i})<19$ (atoms~cm$^{-2}$).
This is a regime where shielding of UV photons becomes extremely difficult even in the
presence of dust. However, it is possible that such systems are missed
in our search. Indeed, as seen in Fig.~\ref{fig:nhi}, there is a trend
for neutral atomic-hydrogen column density to increase with \ion{C}{i}
equivalent width. The gradually decreasing completeness fraction of
the survey below $W_\mathrm{r}(\lambda 1560)\approx 0.4$~\AA\ (see
Sect.~\ref{sec:identification}) would therefore preclude low
$N(\ion{H}{i})$ systems from appearing in our sample.

Above the DLA limit, where the incompleteness fraction of our survey
is less of an issue, it appears that \ion{C}{i}-selected absorbers
do not follow the statistics of \ion{H}{i}-selected DLAs. 
Although the number of \ion{C}{i} systems with $\log N(\ion{H}{i})>20.3$
(atoms~cm$^{-2}$) is small (i.e., only 9 systems), it is apparent that
the overall $N(\ion{H}{i})$ distribution of \ion{C}{i} systems is relatively
flat. A two-sided Kolmogorov-Smirnov test applied to all absorbers
with $\log N(\ion{H}{i})>20.3$ (atoms~cm$^{-2}$) gives a probability of
only 17\% that the
two distributions come from the same parent population (see inset of
Fig.~\ref{fig:histHI}). This could be explained by a larger number of
velocity components in higher \ion{H}{i} column-density gas, thereby increasing the
probability of detecting \ion{C}{i}. Moreover, large amounts of
shielded gas are probably the consequence of the line-of-sight passing
through the absorbing galaxy at small impact parameter, in which case
we can expect the $N(\ion{H}{i})$ distribution to be flatter than that of
the overall DLA population.

%{\bf PPJ: Ya aussi une autre raison qu'il faudrait signaler: c'est
%  qu'on manque peut etre les systemes a disons logNHI=21 et fort
%  contenu en poussières; du coup le quasars est sous la limite
%  SDSS... Faudrait le signaler. Ce n'est pas logNHI qui est en cause
%  mais logNHIxContenuenpoussiere.}

The strongest DLA found among the \ion{C}{i} absorbers in the \ion{H}{i}
sub-sample has $N(\ion{H}{i})=10^{21.8}$ atoms cm$^{-2}$. It is
however located at $z_{\rm abs}\approx z_{\rm em}$. Even ignoring
proximate systems, it is yet
surprising that three intervening DLAs with $N(\ion{H}{i})\ge 10^{21}$
atoms cm$^{-2}$ are present in such a small absorber sample. From DLA
statistics alone, the probability of randomly selecting three DLAs
that strong out of a sample of six DLAs is only 6\%.
%PN: fraction of >21 is 17% then Bernouilli of 3 sucess out of 6 with P=0.176 --> 6%
% For the latter systems, the number statistics is low as only four of
% them have measured $N(\ion{H}{i})$. It seems however that these could
% have somewhat (i.e., typically by a factor of four) larger
% \ion{H}{i} contents than the intervening \ion{C}{i} absorbers.
There is therefore a probable excess of strong DLAs among \ion{C}{i}
absorbers. While the dust content of these systems is significant (see
Sects.~\ref{sec:colours} and \ref{sec:ebv}),
their dust-to-gas ratio must be limited. %Strong DLAs
%in general are expected to contain large amounts of neutral gas and
%possibly cold gas so that \ion{C}{i} lines should be easy to
%detect. 
Indeed, dust reddening and extinction of the background QSOs will
inevitably reduce the incidence of strong and dusty DLAs in
magnitude-limited QSO samples. This implies that the actual proportion of
strong DLAs among \ion{C}{i} systems in general is likely to be even
higher than what we here found out.

% This must be related to extinction. Conclude in Sect. 8 that the
% shape of the N(HI) distribution of CI systems is the result of both
% detection limit / gas self-shielding and CI/HI / extinction of the
% background source increasing with N(HI). Note that CI systems do not
% need to follow the N(HI) distribution of DLAs unless CI/HI is a
% constant, but the latter is probably not the case.

\section{Empirical relations in the sample}

\label{sec:relations}

Based on the results presented in the previous sections, we now study the
existence of empirical relations between the different quantities
measured in this work: neutral atomic-carbon and neutral
atomic-hydrogen contents, the reddening \ion{C}{i}-selected absorbers
induce on their background QSOs and the strength of possible 2175~\AA\
extinction features. Because the \ion{C}{i}\,$\lambda$1560 transition
line is weaker and hence exhibits less saturation than
\ion{C}{i}\,$\lambda$1656, we adopt the equivalent width of the former
as a proxy for the amount of neutral atomic carbon in the systems.

\begin{figure}[htb]
%\centering{\hbox{
%\psfig{figure=NHI.ps,width=8.9cm,clip=,bbllx=25.pt,bblly=115.pt,bburx=543.pt,bbury=704.pt,angle=90.}}}
\resizebox{\hsize}{!}{\includegraphics[angle=90.]{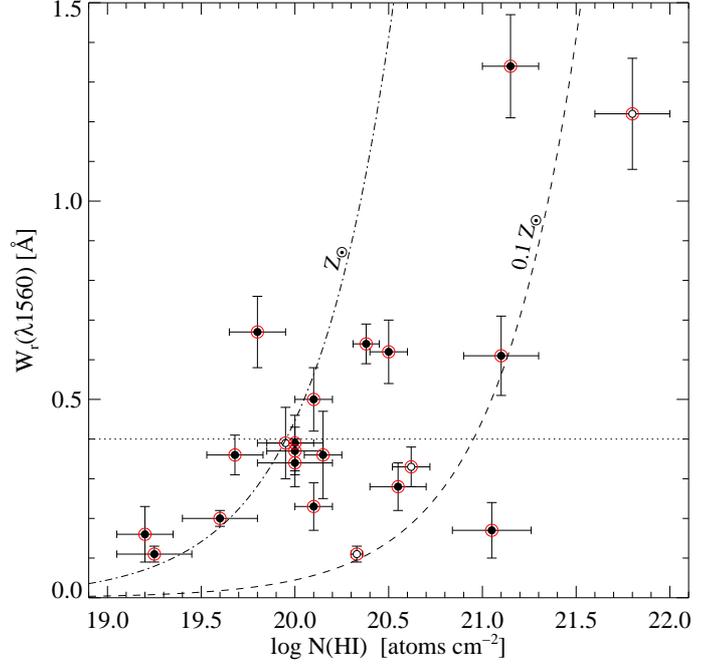}}
\caption{\ion{C}{i}\,$\lambda$1560 rest-frame equivalent width versus
  $N(\ion{H}{i})$, the neutral atomic-hydrogen column density of \ion{C}{i}
  absorbers. Intervening (resp. proximate) systems are displayed as
  filled (resp. empty) black circles. Data points are encircled in red as they
  correspond to the sub-sample of \ion{C}{i} systems with measured $N(\ion{H}{i})$, i.e.,
  the \ion{H}{i} sub-sample. The dashed (resp. dashed-dotted) curve
  materializes the limit above which systems are expected to have in excess
  of one-tenth of Solar (resp. in excess of Solar) metallicity (see text). The
  completeness limit of the survey is indicated by the horizontal line.\label{fig:nhi}}
\end{figure}

In Fig.~\ref{fig:nhi}, we plot the \ion{C}{i}\,$\lambda$1560
rest-frame equivalent width versus $\log N(\ion{H}{i})$ for the
\ion{C}{i} absorbers from the \ion{H}{i} sub-sample. Both quantities appear
to be weakly correlated. A Kendall rank-correlation test indicates
the significance of the correlation to be $1.8\sigma$ only. There is
therefore a tendency for strong DLAs to have larger \ion{C}{i}
equivalent widths but at the same time, for a given \ion{H}{i} column
density, the \ion{C}{i} content can vary substantially from one system
to another. Large values of $W_\mathrm{r}(\lambda 1560)$ are
observed in DLAs but also in sub-DLAs. The fraction of shielded and
probably cold gas could actually be large in some of these sub-DLAs.
From the optically-thin approximation applied to the
\ion{C}{i}\,$\lambda$1560 absorption line and assuming the ionization
equilibrium relation, $N($\ion{C}{i}$)/N($\ion{C}{ii}$)\sim 0.01$,
valid for the cold neutral medium \citep[CNM; see,
e.g.,][]{2011ApJ...734...65J}, a lower limit on the gas metallicity can be
derived:
\begin{equation}\label{eq:met}
  [\mathrm{X}/\mathrm{H}]\ga 18.35+\log\left(\frac{W_\mathrm{r}(\lambda 1560)}{0.01 \times N(\ion{H}{i})}\right)
\end{equation}
For $W_\mathrm{r}(\lambda 1560)=0.4$~\AA\ and $\log N(\ion{H}{i})=20$
(atoms~cm$^{-2}$), the metallicity should be of the order of Solar. More generally,
the dashed and dashed-dotted curves in Fig.~\ref{fig:nhi} were calculated using
the above equation assuming metallicities of one-tenth of Solar and Solar respectively.
Within measurement uncertainties, most of the \ion{C}{i} systems
lie in between these two curves. If the medium
probed by the line-of-sight is a mixture of cold and warm gas, the
metallicity of the systems will be even higher. However, if part of the hydrogen is
in molecular form, the metallicity will be lower. For
the whole \ion{C}{i}-absorber sample, Eq.~\ref{eq:met} implies a
metallicity distribution ranging between [X/H$]=-1.4$ and
metallicities in excess of Solar, with a median value of [X/H$]\approx
-0.5$. This means that the metallicities of \ion{C}{i} absorbers would
on average be at least ten times larger than those of typical
DLAs (for the latter, see, e.g., \citet{2012ApJ...755...89R}). This should be confirmed by accurate measurements
of metal column densities.

\begin{figure}[htb]
%\centering{\hbox{
%\psfig{figure=NHI2.ps,width=8.9cm,clip=,bbllx=25.pt,bblly=115.pt,bburx=543.pt,bbury=704.pt,angle=90.}}}
\resizebox{\hsize}{!}{\includegraphics[angle=90.]{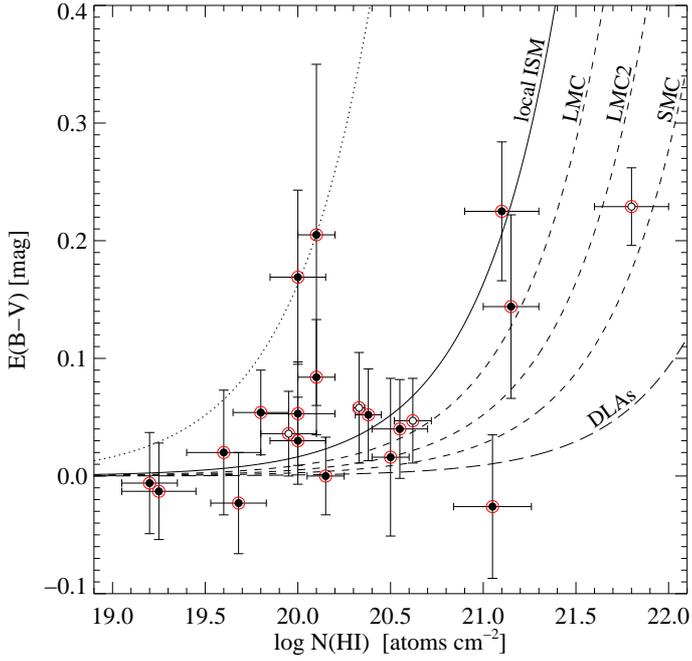}}
\caption{Reddening of the background QSOs with detected \ion{C}{i}
  absorption versus $N(\ion{H}{i})$ of the \ion{C}{i} absorbers.
  Symbol conventions are the same as in Fig.~\ref{fig:nhi}. The solid line shows
  the relation observed in the local ISM where $E($B-V$)/N($H$)=1.63\times 10^{-22}$
  mag atoms$^{-1}$ cm$^2$ \citep{2012ApJS..199....8G}. The dotted line
  illustrates a reddening per hydrogen atom ten times higher than that.
  The dashed curves correspond to the observations of the Magellanic
  Clouds \citep{2003ApJ...594..279G}. The long-dashed line shows the
  relation derived for typical DLAs at $z_\mathrm{abs}\approx 2.8$
  \citep[e.g.,][]{2008A&A...478..701V,2012MNRAS.419.1028K}.\label{fig:nhi2}}
\end{figure}

In Fig.~\ref{fig:nhi2}, we display the relation between $E($B-V$)$
and $\log N(\ion{H}{i})$ among the \ion{C}{i} absorbers from the \ion{H}{i}
sub-sample. Here again, the data points are highly scattered. Most of
the systems are associated with low albeit consistently non-zero QSO
reddening. Since most of the $N(\ion{H}{i})$ values are relatively low,
the measured amounts of reddening, with median $E($B-V$)\sim 0.045$, are actually
remarkable. This departs from what is observed in the overall DLA population
where the reddening is usually negligible \citep[see,
e.g.,][]{2008A&A...478..701V,2012MNRAS.419.1028K}. The latter authors
have shown that DLAs at $z_\mathrm{abs}\approx 2.8$ typically induce a
reddening $E($B-V$)\sim 5\times 10^{-3}$~mag. Apart from a few
outliers, most of the \ion{C}{i} systems have reddening properties
consistent with those of the Galactic ISM. This is represented in
Fig.~\ref{fig:nhi2} by the solid line. In the Galaxy, the
reddening induced along a line-of-sight is indeed directly proportional to
the neutral atomic-hydrogen column density, with
$E($B-V$)/N($H$)=1.63\times 10^{-22}$ mag atoms$^{-1}$ cm$^2$
\citep{2012ApJS..199....8G}. Only two \ion{C}{i} systems with large
$N(\ion{H}{i})$ are more consistent with what is seen in typical DLAs and/or
along SMC lines-of-sight, where the above ratio is smaller than in the Galaxy.
Two other \ion{C}{i} systems with low
$N(\ion{H}{i})$ might also deviate from the Galactic relation being
consistent with a ten-times larger ratio. We caution however that the
uncertainties on the reddening measurements are fairly large. If real, this
would imply in the latter systems the existence of a grain chemistry more evolved
than in the Galaxy with a larger fraction of big grains over very small grains
\citep[e.g.,][]{1992ApJ...395..130P}. This is opposite to the trend
observed in the Magellanic Clouds. In such systems, strong
2175~\AA\ absorption is expected. Interestingly, this is what is observed in practice as
these two \ion{C}{i} systems exhibit two of the strongest three $A_{\rm bump}$ values of
the \ion{H}{i} sub-sample.

\begin{figure*}[htb]
%\centering{\hbox{
%\psfig{figure=EBV.ps,width=9.1cm,clip=,bbllx=25.pt,bblly=151.pt,bburx=543.pt,bbury=704.pt,angle=90.}
%\psfig{figure=Abump.ps,width=9.1cm,clip=,bbllx=25.pt,bblly=151.pt,bburx=543.pt,bbury=704.pt,angle=90.}}}
\resizebox{\hsize}{!}{\includegraphics[angle=90.]{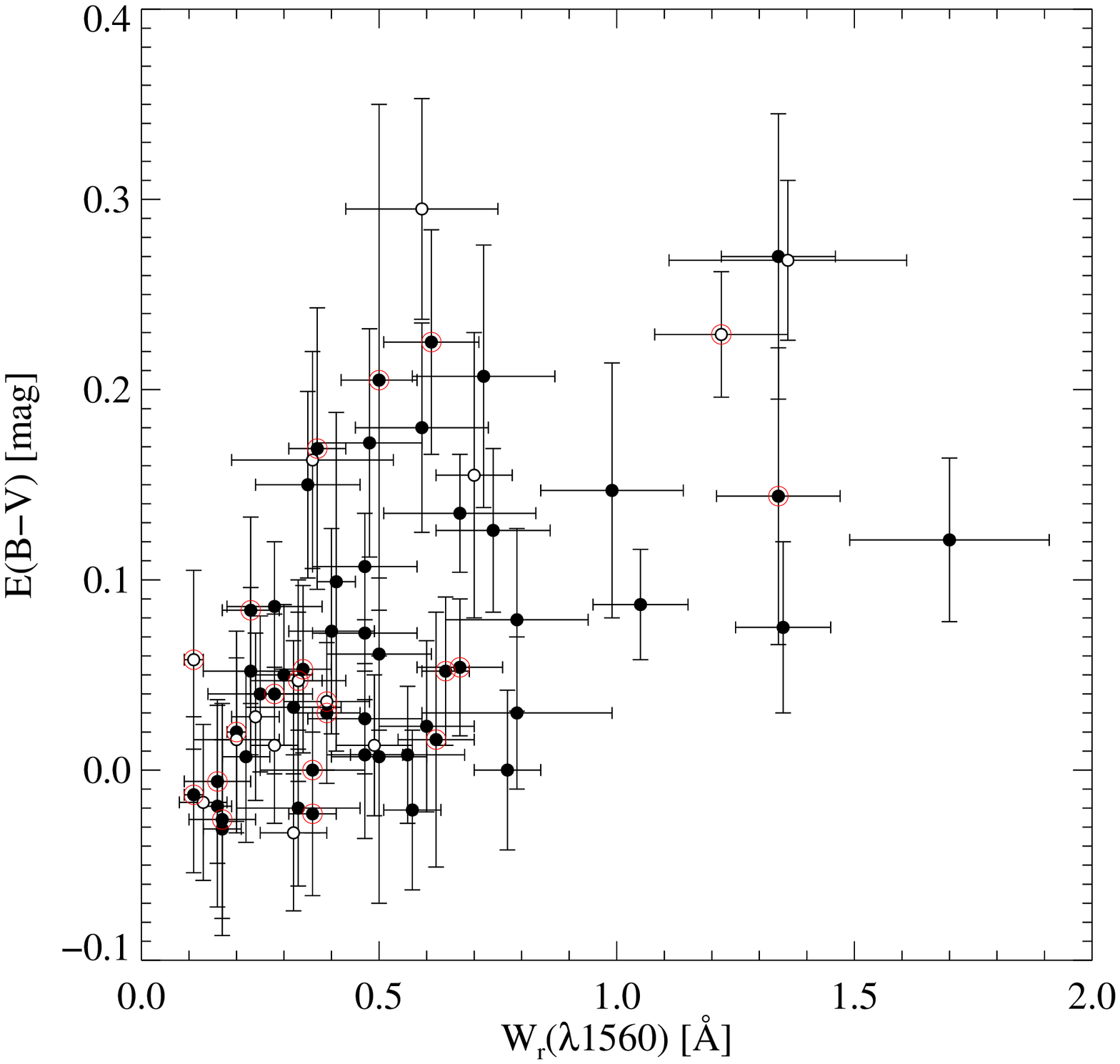}\includegraphics[angle=90.]{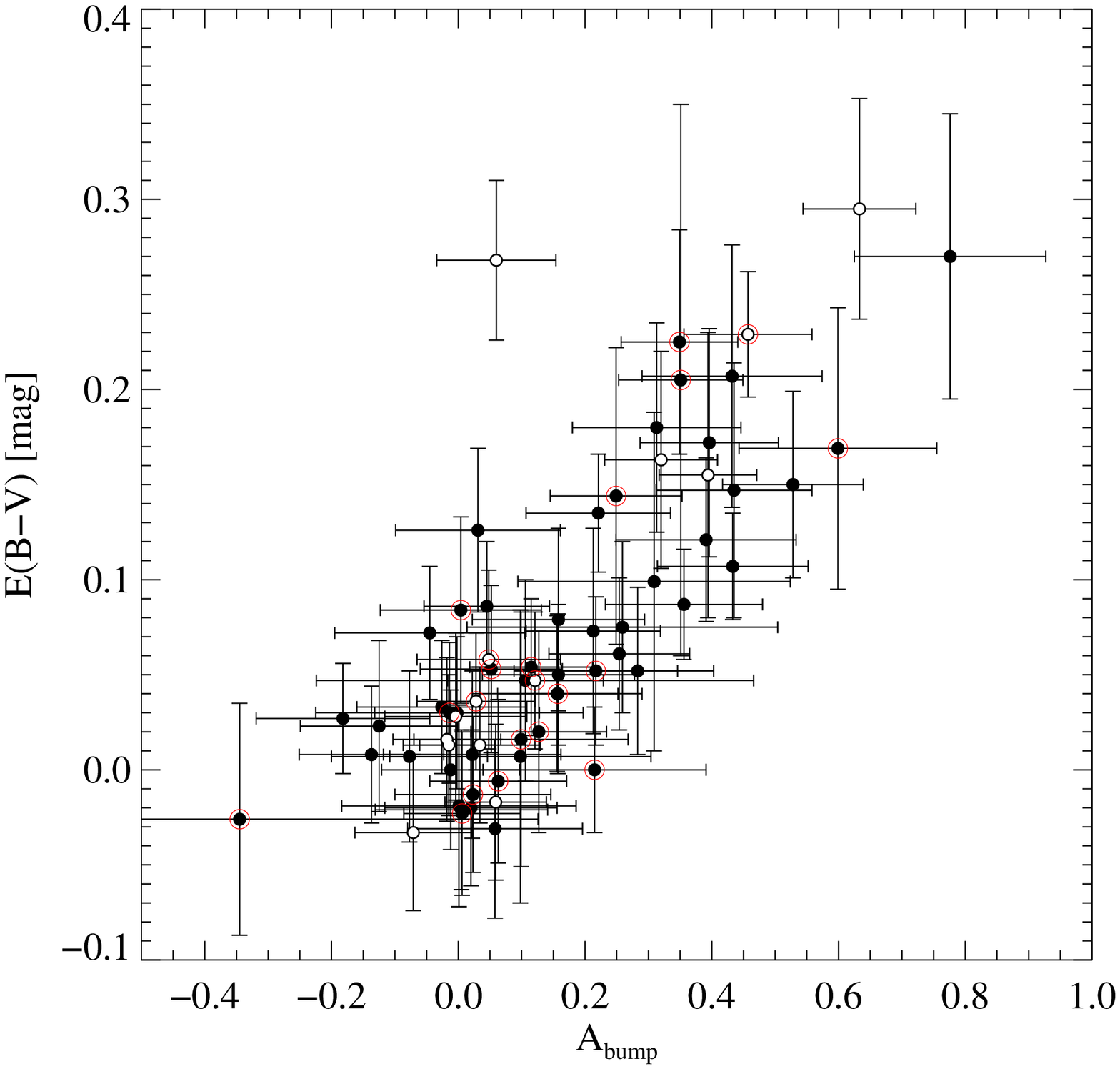}}
\caption{{\it Left panel:} Reddening of the background QSOs with
  detected \ion{C}{i} absorption (overall sample) versus
  \ion{C}{i}\,$\lambda$1560 rest-frame equivalent width. Symbol
  conventions are the same as in Fig.~\ref{fig:nhi}. Data points encircled in red correspond to
  the \ion{H}{i} sub-sample. {\it Right
    panel:} Same as in the left panel with the exception that reddening is plotted
  against 2175~\AA\ bump strength.\label{fig:ebv}}
\end{figure*}

To investigate the characteristics of the \ion{C}{i} absorbers further, we look
in Fig.~\ref{fig:ebv} in more detail at the properties of dust in these systems. In the
left panel of this figure, $E($B-V$)$ and $W_\mathrm{r}(\lambda 1560)$ are
found to be correlated with each other at the $4.4\sigma$ significance level. This
is noteworthy as different degrees of saturation of the \ion{C}{i}\,$\lambda$1560 line
are expected to produce scatter in this relation. This implies that the neutral-carbon
content of the \ion{C}{i} systems is intimately related to the reddening induced along
the line-of-sight or, equivalently, that the amounts of shielded gas and dust are tightly
inter-connected. We also note that two of the largest three $E($B-V$)$ values in this
plot correspond to systems located at $z_{\rm abs}\approx z_{\rm em}$. This may
however be the result of small number statistics as the reddening induced by the other
proximate systems in our sample varies substantially from one system to another.

The relation between $E($B-V$)$ and the UV bump strength, $A_\mathrm{bump}$,
which we previously determined independently from $E($B-V$)$ (see Sect.~\ref{sec:abump}),
is shown in the right panel of Fig.~\ref{fig:ebv}. It can be seen that both quantities
are tightly correlated ($6.0\sigma$). A linear least-squares fit (linear correlation
coefficient of $r=0.77$), taking into account errors in both parameters, gives:
$E($B-V$)\simeq 0.43\times A_\mathrm{bump}$. The 2175~\AA\ extinction feature
is detected at more than $2\sigma$ (95\% confidence level) in about 30\% of
the \ion{C}{i} systems. In such cases, we find $A_\mathrm{bump}\sim 0.4$
and $E($B-V$)\sim 0.2$~mag or, equivalently, $A_{\rm V}\sim 0.6$~mag. These
values are comparable to what \citet{2011MNRAS.416.1871B} have found when
targeting the strongest \ion{Mg}{ii} systems from SDSS DR\,6
[$1<W_\mathrm{r}(\lambda 2796)<5$~\AA] where the 2175~\AA\ absorption is
detected on a statistical basis only \citep[see also][for candidate 2175~\AA\ absorption
in similar systems]{2011ApJ...732..110J}. Interestingly, our measured UV bump
strengths are also comparable to what has been observed along GRB lines-of-sight
at similarly low levels of extinction, e.g., towards GRB\,080605 \citep[see][]{2012ApJ...753...82Z}.

In the local Universe, 2175~\AA\ absorption can be observed together
with reddening values as low as $\sim 0.2$~mag \citep[see,
e.g.,][]{2007ApJ...663..320F}. Even at such low levels of
reddening, the UV bump is significantly stronger along Galactic
lines-of-sight, i.e., by up to a factor of ten, than in the present
\ion{C}{i} absorber sample and/or through GRB host galaxies. This
discrepancy may be explained if these high-redshift systems probe regions
of the ISM affected by a star formation more vigorous than in the Galaxy.
A similar argument was proposed by \citet{2003ApJ...594..279G}
with the aim of explaining the variety of LMC and SMC extinction curves.
In fact, most of the \ion{C}{i} absorbers at the high end of the reddening tail
in our sample are best-fit using the extinction law of the LMC2 super-shell
near the 30~Dor star-forming region\footnote{Super-giant shells with
  sizes approaching 1~kpc form the largest structures seen in the ISM
  of galaxies where large amounts of kinetic energy are contributed by
  multiple supernovae explosions and energetic stellar winds.} (see
left panel of Fig.~\ref{fig:histo}). This means that the far-UV rise
of the extinction curve is enhanced and the carriers of the 2175~\AA\
absorption are depleted compared to Galactic lines-of-sight. This is
probably the consequence of a high UV flux and/or the mechanical
feedback from stars \citep[e.g.,][]{2011A&A...533A.117F} in the
vicinity of the \ion{C}{i} systems.
% In contrast, typical DLAs can be seen as quiescent systems
% associated with low and/or episodic star-formation activity, in
% accord with their low metallicities.
In contrast, the lack of a UV bump in typical DLAs
\citep[e.g.,][]{2012MNRAS.419.1028K} is probably
intrinsic to their low dust and metal contents as the lines-of-sight are
likely to pass at large impact parameters from the absorbing galaxy.

The main outlier in the right panel of Fig.~\ref{fig:ebv}, which exhibits
high reddening but no UV bump, is a proximate system. This
is consistent with the above picture where the enhanced UV radiation
field from the QSO and/or star-forming regions within the QSO host
galaxy are expected to deplete the carriers of the 2175~\AA\ absorption.

% dust composition and evolution: graphite+silicate (LMC2, MW)
% vs. silicate (SMC): graphite and silicate coexist mainly during the
% early phases of galaxy chemical evolution (at most 2~Myrs), then
% graphites are destroyed and not replaced. See also Updike et al.
% \citet{2012ApJ...753...82Z}

\section{Conclusions}

\label{sec:conclusions}

In this work, we presented a new population of QSO absorbers selected
directly from the properties of the shielded gas, namely the strongest
\ion{C}{i} absorbers, detected in low-resolution QSO spectra from the
SDSS-II DR\,7 database. These \ion{C}{i} absorbers, with
$W_\mathrm{r}(\lambda 1560)\ge 0.4$~\AA, are more than one hundred-times rarer
than DLAs at $z_\mathrm{abs}=2.5$. Their number per unit redshift is
increasing significantly below $z_\mathrm{abs}=2$, probably coupled to an increase
in the star-formation efficiency at these
redshifts. \citet{2012A&A...544A..21G} reported a similarly
high detection rate of 21-cm absorbers towards even lower redshifts
among strong \ion{Mg}{ii} systems, which they argued must be related
to the evolution of the CNM filling factor in the latter absorbers.

%2) mettre du poids en Section 8 sur l'evolution en redshift du gas froid
%riche en metaux et l'origine
%des systemes CI (disques??).

The \ion{H}{i} column-density distribution of \ion{C}{i}-selected absorbers
is flatter than that of \ion{H}{i}-selected absorbers. While sub-DLAs
have much larger cross-section than DLAs, this can be understood as
the shielding of the gas is more difficult at low \ion{H}{i} column densities and
the number of clouds along the line-of-sight is probably
smaller. Cold and dusty gas as traced by \ion{C}{i} absorbers is also
more likely to be found at small impact parameters from the absorbing
galaxies where a flatter $N(\ion{H}{i})$ distribution is expected. Indeed,
despite a likely bias against strong DLAs with large amounts of dust,
we find there is among \ion{C}{i} systems a probable excess of strong DLAs
with $\log N(\ion{H}{i})>21$ (atoms~cm$^{-2}$) compared to systematic
DLA searches. This is
reminiscent of the $N(\ion{H}{i})$ distribution of DLAs within GRB host
galaxies which is skewed towards extremely strong DLAs
\citep[see fig.~10 in][]{2009ApJS..185..526F}.

The reddening and therefore the presence of dust along the QSO
lines-of-sight with detected \ion{C}{i} absorption is directly related
to the amount of shielded gas but depends weakly on the total \ion{H}{i}
column density. The latter can indeed vary by more than a factor of ten for
the same \ion{C}{i} rest-frame equivalent width. This is probably the
consequence of the shielded gas being clumpy while \ion{H}{i}
absorption samples simultaneously warm diffuse neutral clouds and
cold, high-metallicity dusty pockets of gas. The presence of dust
inducing significant reddening of the background QSOs and/or 2175~\AA\
extinction features are ubiquitous in about 30\% of the \ion{C}{i}
absorbers. Several systems like these have been found before
\citep[see, e.g.,][]{2008MNRAS.391L..69S,2012ApJ...760...42W}. Here, we
find that the UV bump is weak compared to Galactic lines-of-sight
exhibiting the same amount of reddening. We interpret this as being
the consequence of star formation in the vicinity of the systems.

It is likely that the metal and molecular contents of \ion{C}{i}
absorbers are high and actually higher than those of most DLAs studied
till now. High-resolution spectroscopic follow-up observations of the
present sample therefore opens up the door to systematic searches for
carbon monoxide \citep[CO; see][]{2011A&A...526L...7N} and molecules
like CN and CH as well as diffuse interstellar bands at high
redshift. Such a spectroscopic campaign will be presented in a
companion paper. The typical reddening induced by \ion{C}{i} absorbers
along with the relation between reddening and shielded-gas column
density imply that the extinction could be high in some DLAs with
\ion{C}{i} absorption. If
strong dusty DLAs exist, they probably have been missed in the current
magnitude-limited QSO samples
\citep[see also][]{1998A&A...333..841B,2008A&A...478..701V}.
Some of the QSO lines-of-sight
identified here, as well as those which may be found by extending the
present survey to even larger databases\footnote{Note that from the
  Baryon Oscillation Spectroscopic Survey, which is part of SDSS-III,
  relatively few additional \ion{C}{i} systems are expected since the bulk
  of the new QSOs is at $z_\mathrm{em}\sim 3$, which provides
  shorter \ion{C}{i}-absorption path length. 
%PN: J'ai enlevé cette remarque car justement, les qsos faibles pourraient 
%avoir plus de chance de passer dans CI. De plus, elle n'est pas necessaire.
%On pourrait meme enlever la remarque en entier, d'ailleurs, car eBOSS 
%vient a la rescousse. 
%, and (ii) the
%  QSOs are generally fainter and the S/N ratio of the spectra is lower
%  than in SDSS-II DR\,7 \citep[see][]{2012A&A...548A..66P}.
}, will result in exceedingly long integration times on
high-resolution spectrographs installed on 8-10\,m class
telescopes. These will however be targets of choice for the coming
generation of Extremely Large Telescopes.

\begin{acknowledgements}

  PN acknowledges support from the ESO Chile visiting scientist programme.
  RS and PPJ gratefully acknowledge support from the
  Indo-French Centre for the Promotion of Advanced Research (Centre
  Franco-Indien pour la Promotion de la Recherche Avanc\'ee) under
  contract No.~4304-2.
  The authors of this paper also acknowledge the tremendous effort put
  forth by the Sloan Digital Sky Survey team to produce and release
  the SDSS survey. Funding for SDSS and SDSS-II has been provided
  by the Alfred P. Sloan Foundation, the Participating Institutions,
  the National Science Foundation, the U.S. Department of Energy, the
  National Aeronautics and Space Administration, the Japanese
  Monbukagakusho, the Max Planck Society, and the Higher Education
  Funding Council for England. The SDSS Web Site is
  http://www.sdss.org/. The SDSS is managed by the Astrophysical
  Research Consortium for the Participating Institutions. The
  Participating Institutions are the American Museum of Natural
  History, Astrophysical Institute Potsdam, University of Basel,
  University of Cambridge, Case Western Reserve University, University
  of Chicago, Drexel University, Fermilab, the Institute for Advanced
  Study, the Japan Participation Group, Johns Hopkins University, the
  Joint Institute for Nuclear Astrophysics, the Kavli Institute for
  Particle Astrophysics and Cosmology, the Korean Scientist Group, the
  Chinese Academy of Sciences (LAMOST), Los Alamos National
  Laboratory, the Max-Planck-Institute for Astronomy (MPIA), the
  Max-Planck-Institute for Astrophysics (MPA), New Mexico State
  University, Ohio State University, University of Pittsburgh,
  University of Portsmouth, Princeton University, the United States
  Naval Observatory, the University of Washington.

\end{acknowledgements}

\bibliographystyle{aa}
\bibliography{cico}

%------------------------------------------------------------------------------

\clearpage

%------------------------------------------------------------------------------

%------------------------------------------------------------------------------

%------------------------------------------------------------------------------

%------------------------------------------------------------------------------

\clearpage

%------------------------------------------------------------------------------

\clearpage

%------------------------------------------------------------------------------

\clearpage

%------------------------------------------------------------------------------

%------------------------------------------------------------------------------

%------------------------------------------------------------------------------

%------------------------------------------------------------------------------

%--------------------------------------------------------------------------------

\clearpage

\begin{table*}[htb]
  \caption{\ion{C}{i} systems identified in SDSS-DR7 QSO spectra.\label{tab:ci}}
\begin{tabular}{ccccccccccc}
\hline
\hline
QSO name & plate & MJD & fibre &
$z_\mathrm{em}$  &
$z_\mathrm{abs}$ &
$W_\mathrm{r}(\lambda 1560)^\mathrm{a}$ &
$W_\mathrm{r}(\lambda 1656)^\mathrm{a}$ & S/N$^\mathrm{b}$ & $\log N(\ion{H}{i})^\mathrm{c}_\mathrm{SDSS}$ & $\log N(\ion{H}{i})^\mathrm{d}_\mathrm{UVES}$\\
& & & & & & [\AA ] & [\AA ] & & [atoms~cm$^{-2}$] & [atoms~cm$^{-2}$]\\
\hline
J\,021606.13$-$002104.9 & 0405 & 51816 & 114 & 2.22 & 1.737 & $0.33\pm 0.10$ & $0.77\pm 0.08$ &           14 &       ...               &       ...               \\%XS
J\,030049.24$-$072137.8 & 0458 & 51929 & 437 & 2.11 & 1.536 & $0.79\pm 0.15$ & $1.17\pm 0.21$ & \phantom{1}5 &       ...               &       ...               \\%UVES
J\,080801.74$+$330009.2 & 0825 & 52289 & 166 & 1.90 & 1.888 & $0.24\pm 0.05$ & $0.39\pm 0.05$ &           18 &       ...               &       ...               \\%N
J\,081116.06$+$083837.7 & 2570 & 54081 & 566 & 2.07 & 1.906 & $0.23\pm 0.06$ & $0.21\pm 0.06$ &           15 &       ...               & $20.10\pm 0.10$         \\%UVES
J\,081540.60$+$264021.6 & 1266 & 52709 & 355 & 1.94 & 1.681 & $1.70\pm 0.21$ & $1.33\pm 0.16$ & \phantom{1}5 &       ...               &       ...               \\%XS?
J\,082003.40$+$155932.9 & 2272 & 53713 & 378 & 1.95 & 1.547 & $0.72\pm 0.15$ & $0.77\pm 0.11$ & \phantom{1}7 &       ...               &       ...               \\%UVES
J\,085206.65$+$193548.4 & 2281 & 53711 & 452 & 1.91 & 1.788 & $0.34\pm 0.06$ & $0.86\pm 0.07$ &           13 &       ...               & $20.00\pm 0.20$         \\%UVES
J\,085437.60$+$031734.8 & 0565 & 52225 & 480 & 2.24 & 1.567 & $0.40\pm 0.09$ & $0.43\pm 0.07$ &           12 &       ...               &       ...               \\%XS
J\,085726.79$+$185524.3 & 2281 & 53711 & 630 & 1.89 & 1.730 & $0.20\pm 0.02$ & $0.43\pm 0.04$ &           33 &       ...               & $19.60\pm 0.20$         \\%UVES - CO
J\,090558.75$+$553039.2 & 0450 & 51908 & 470 & 1.57 & 1.579 & $0.13\pm 0.05$ & $0.15\pm 0.05$ &           16 &       ...               &       ...               \\%N
J\,090942.56$+$532126.5 & 0553 & 51999 & 425 & 2.07 & 1.628 & $0.47\pm 0.08$ & $0.70\pm 0.09$ &           12 &       ...               &       ...               \\%N
J\,091516.27$+$071750.9 & 1194 & 52703 & 506 & 2.38 & 1.681 & $0.50\pm 0.10$ & $0.95\pm 0.13$ &           11 &       ...               &       ...               \\%Forest
J\,091721.37$+$015448.1 & 0473 & 51929 & 426 & 2.18 & 2.107 & $0.67\pm 0.16$ & $0.93\pm 0.16$ & \phantom{1}5 &       ...               &       ...               \\%XS*
J\,092759.79$+$154321.8 & 2579 & 54068 & 364 & 1.81 & 1.731 & $0.61\pm 0.10$ & $0.69\pm 0.09$ & \phantom{1}9 &       ...               & $21.10\pm 0.20$         \\%UVES
J\,094529.37$+$552525.7 & 0556 & 51991 & 045 & 2.24 & 1.867 & $0.74\pm 0.12$ & $0.91\pm 0.13$ & \phantom{1}7 &       ...               &       ...               \\%N
J\,095114.97$+$545736.5 & 0769 & 52282 & 458 & 1.80 & 1.613 & $0.77\pm 0.07$ & $0.78\pm 0.06$ &           14 &       ...               &       ...               \\%N
J\,101544.34$+$310617.2 & 1953 & 53358 & 604 & 1.56 & 1.596 & $0.20\pm 0.09$ & $0.34\pm 0.09$ & \phantom{1}9 &       ...               &       ...               \\%N
J\,104705.76$+$205734.5 & 2478 & 54097 & 558 & 2.01 & 1.775 & $0.59\pm 0.14$ & $0.96\pm 0.13$ & \phantom{1}7 &       ...               &       ...               \\%UVES/XS* - CO
J\,105436.96$+$542343.7 & 0907 & 52373 & 129 & 1.50 & 1.513 & $0.49\pm 0.08$ & $0.87\pm 0.08$ &           10 &       ...               &       ...               \\%N
J\,105746.42$+$662250.5 & 0490 & 51929 & 179 & 2.34 & 2.088 & $0.30\pm 0.06$ & $0.48\pm 0.06$ &           13 &       ...               &       ...               \\%N
J\,111756.53$+$143716.9 & 1753 & 53383 & 353 & 2.09 & 2.001 & $0.67\pm 0.09$ & $0.65\pm 0.08$ &           10 &       ...               & $19.80\pm 0.15$         \\%UVES
J\,112250.95$+$143732.5 & 1753 & 53383 & 477 & 2.05 & 1.554 & $0.17\pm 0.04$ & $0.19\pm 0.03$ &           28 &       ...               &       ...               \\%XS?
J\,112913.29$-$023740.9 & 0327 & 52294 & 308 & 1.86 & 1.623 & $0.60\pm 0.10$ & $0.99\pm 0.12$ & \phantom{1}8 &       ...               &       ...               \\%UVES
J\,113341.29$-$005740.1 & 0282 & 51658 & 215 & 1.68 & 1.706 & $1.36\pm 0.25$ & $1.24\pm 0.24$ & \phantom{1}4 &       ...               &       ...               \\%XS*
J\,114141.84$+$444206.1 & 1367 & 53083 & 499 & 1.96 & 1.903 & $0.48\pm 0.11$ & $0.76\pm 0.13$ & \phantom{1}7 &       ...               &       ...               \\%N
J\,115153.87$+$150945.0 & 1762 & 53415 & 495 & 3.05 & 2.400 & $0.36\pm 0.11$ & $0.37\pm 0.11$ & \phantom{1}7 & \phantom{$^\mathrm{e}$}$20.15\pm 0.33^\mathrm{e}$ & $20.15\pm 0.10$         \\%XS?
J\,115705.51$+$615521.7 & 0777 & 52320 & 107 & 2.51 & 2.460 & $1.22\pm 0.14$ & $1.30\pm 0.14$ & \phantom{1}7 & \phantom{$^\mathrm{f}$}$21.80\pm 0.20^\mathrm{f}$ &       ...                \\%N
J\,120935.79$+$671715.7 & 0493 & 51957 & 178 & 2.03 & 1.843 & $0.47\pm 0.11$ & $0.48\pm 0.11$ & \phantom{1}8 &       ...               &       ...               \\%N
J\,122825.67$+$303038.6 & 2234 & 53823 & 430 & 3.91 & 3.098 & $0.17\pm 0.07$ & $0.26\pm 0.07$ &           11 & \phantom{$^\mathrm{e}$}$21.05\pm 0.21^\mathrm{e}$ &       ...               \\%N
J\,123714.61$+$064759.6 & 1628 & 53474 & 193 & 2.78 & 2.691 & $0.37\pm 0.06$ & $0.77\pm 0.08$ &           14 & \phantom{$^\mathrm{e}$}$20.15\pm 0.28^\mathrm{e}$ & \phantom{$^\mathrm{g}$}$20.00\pm 0.15^\mathrm{g}$ \\%UVES/XS* - CO published
J\,124708.43$+$500320.8 & 1278 & 52735 & 039 & 2.27 & 2.135 & $1.05\pm 0.10$ & $1.42\pm 0.11$ & \phantom{1}9 &       ...               &       ...               \\%N
J\,124829.71$+$284858.1 & 2239 & 53726 & 219 & 1.54 & 1.513 & $0.32\pm 0.07$ & $0.65\pm 0.07$ &           12 &       ...               &       ...               \\%XS?
J\,124841.45$+$302433.0 & 2239 & 53726 & 469 & 2.06 & 1.691 & $0.57\pm 0.06$ & $0.87\pm 0.06$ &           15 &       ...               &       ...               \\%N
J\,125552.60$+$223424.4 & 2649 & 54212 & 507 & 1.82 & 1.526 & $0.25\pm 0.11$ & $1.18\pm 0.28$ & \phantom{1}8 &       ...               &       ...               \\%XS*
J\,130225.28$+$211158.6 & 2650 & 54505 & 296 & 1.76 & 1.656 & $0.56\pm 0.12$ & $0.49\pm 0.10$ & \phantom{1}8 &       ...               &       ...               \\%XS?
J\,130628.87$+$281550.8 & 2009 & 53904 & 288 & 2.10 & 2.012 & $0.28\pm 0.10$ & $0.56\pm 0.10$ & \phantom{1}8 &       ...               &       ...               \\%UVES
J\,130828.43$+$584000.6 & 0958 & 52410 & 224 & 3.09 & 2.473 & $0.39\pm 0.07$ & $0.61\pm 0.14$ &           11 & $20.00\pm 0.15$         &       ...               \\%N
J\,131129.11$+$222552.6 & 2651 & 54507 & 398 & 3.14 & 3.093 & $0.33\pm 0.05$ & $0.51\pm 0.08$ &           16 & $20.50\pm 0.10$         & $20.62\pm 0.10$         \\%UVES
J\,131400.57$+$054319.5 & 0850 & 52338 & 620 & 1.89 & 1.583 & $0.23\pm 0.10$ & $0.70\pm 0.10$ & \phantom{1}8 &       ...               &       ...               \\%XS?
J\,134122.51$+$185214.0 & 2642 & 54232 & 494 & 2.00 & 1.544 & $0.16\pm 0.03$ & $0.15\pm 0.04$ &           27 &       ...               &       ...               \\%XS?
J\,134601.10$+$064408.4 & 1803 & 54152 & 074 & 2.09 & 1.512 & $0.79\pm 0.20$ & $1.33\pm 0.28$ & \phantom{1}6 &       ...               &       ...               \\%XS?
J\,135122.00$+$461828.5 & 1466 & 53083 & 416 & 1.81 & 1.606 & $0.99\pm 0.15$ & $1.58\pm 0.17$ &           10 &       ...               &       ...               \\%N
J\,141550.47$+$300146.9 & 2129 & 54252 & 288 & 2.08 & 1.676 & $1.35\pm 0.10$ & $1.61\pm 0.12$ &           13 &       ...               &       ...               \\%N
J\,141606.79$+$180403.2 & 2759 & 54534 & 359 & 2.13 & 1.622 & $0.47\pm 0.11$ & $0.80\pm 0.13$ & \phantom{1}8 &       ...               &       ...               \\%
J\,143243.93$+$330746.7 & 1841 & 53491 & 513 & 2.09 & 2.058 & $0.28\pm 0.05$ & $0.44\pm 0.05$ &           17 &       ...               &       ...               \\%N
J\,143657.87$+$291100.6 & 2138 & 53757 & 259 & 1.77 & 1.596 & $0.35\pm 0.11$ & $0.56\pm 0.11$ & \phantom{1}8 &       ...               &       ...               \\%~N
J\,143912.05$+$111740.6 & 1711 & 53535 & 374 & 2.58 & 2.418 & $0.50\pm 0.08$ & $0.68\pm 0.09$ &           10 & \phantom{$^\mathrm{e}$}$20.27\pm 0.24^\mathrm{e}$ & \phantom{$^\mathrm{h}$}$20.10\pm 0.10^\mathrm{h}$ \\%UVES - CO published
J\,144929.27$+$333811.0 & 1646 & 53498 & 065 & 2.17 & 2.021 & $0.47\pm 0.12$ & $0.62\pm 0.12$ & \phantom{1}7 &       ...               &       ...               \\%N
J\,145432.54$+$343523.9 & 1384 & 53121 & 269 & 1.61 & 1.580 & $0.59\pm 0.16$ & $0.59\pm 0.13$ & \phantom{1}6 &       ...               &       ...               \\%N
J\,145953.25$+$012944.2 & 0538 & 52029 & 039 & 1.66 & 1.623 & $0.36\pm 0.17$ & $0.62\pm 0.16$ & \phantom{1}5 &       ...               &       ...               \\%UVES
J\,150738.73$+$415530.6 & 1291 & 52735 & 505 & 1.79 & 1.674 & $0.50\pm 0.11$ & $0.85\pm 0.16$ & \phantom{1}7 &       ...               &       ...               \\%N
J\,152209.12$+$083020.0 & 1721 & 53857 & 192 & 1.93 & 1.627 & $0.32\pm 0.10$ & $0.71\pm 0.11$ & \phantom{1}8 &       ...               &       ...               \\%UVES
J\,160320.76$+$170117.8 & 2200 & 53875 & 318 & 1.99 & 1.890 & $0.16\pm 0.07$ & $0.43\pm 0.08$ &           12 &       ...               & $19.20\pm 0.15$         \\%UVES
J\,160457.51$+$220300.5 & 2205 & 53793 & 328 & 1.98 & 1.641 & $1.34\pm 0.12$ & $1.60\pm 0.13$ & \phantom{1}7 &       ...               &       ...               \\%UVES - CO published
J\,161526.65$+$264813.8 & 1576 & 53496 & 394 & 2.18 & 2.118 & $0.28\pm 0.06$ & $0.56\pm 0.12$ &           15 &       ...               & $20.55\pm 0.15$         \\%UVES
J\,162321.43$+$135532.4 & 2202 & 53566 & 132 & 1.75 & 1.751 & $0.39\pm 0.09$ & $0.60\pm 0.10$ & \phantom{1}9 &       ...               & $19.95\pm 0.15$         \\%UVES
J\,164610.20$+$232923.0 & 1424 & 52912 & 301 & 2.06 & 1.998 & $0.36\pm 0.05$ & $0.58\pm 0.08$ &           15 &       ...               & $19.68\pm 0.15$         \\%
J\,170542.92$+$354340.4 & 0974 & 52427 & 561 & 2.01 & 2.038 & $0.70\pm 0.08$ & $1.40\pm 0.10$ &           10 &       ...               &       ...               \\%UVES - CO
J\,212329.47$-$005053.0 & 0987 & 52523 & 103 & 2.26 & 2.060 & $0.11\pm 0.02$ & $0.11\pm 0.02$ &           41 &       ...               & \phantom{$^\mathrm{i}$}$19.25\pm 0.20^\mathrm{i}$ \\%UVES Murphy
%J\,222910.16$+$141402.2 & 0738 & 52521 & 382 & 2.11 & 1.586 & $0.33\pm 0.13$ & $0.47\pm 0.13$ & \phantom{1}6 &       ...               &       ...               \\%XS?
%J\,225719.04$-$100104.7 & 0724 & 52254 & 150 & 2.08 & 1.836 & $0.64\pm 0.05$ & $0.89\pm 0.05$ &           17 &       ...               & $20.38\pm 0.07$         \\%UVES
%J\,233133.05$-$090246.6 & 0646 & 52523 & 619 & 2.44 & 1.734 & $0.41\pm 0.04$ & $0.35\pm 0.04$ &           20 &       ...               &       ...               \\%
%J\,233156.49$-$090802.0 & 0646 & 52523 & 616 & 2.66 & 2.143 & $1.34\pm 0.13$ & $1.17\pm 0.11$ & \phantom{1}8 &       ...               & $21.15\pm 0.15$         \\%UVES Nestor - CO
%J\,233633.81$-$105841.5 & 0647 & 52553 & 201 & 2.04 & 1.829 & $0.22\pm 0.05$ & $0.40\pm 0.07$ &           15 &       ...               &       ...               \\%XS
%J\,234023.67$-$005327.1 & 0385 & 51877 & 204 & 2.09 & 2.054 & $0.11\pm 0.02$ & $0.17\pm 0.02$ &           37 &       ...               & $20.33\pm 0.03$         \\%UVES HIRES - HD/H2
%J\,235057.87$-$005210.0 & 0386 & 51788 & 137 & 3.03 & 2.426 & $0.62\pm 0.08$ & $0.63\pm 0.08$ &           11 & \phantom{$^\mathrm{e}$}$20.39\pm 0.27^\mathrm{e}$ & \phantom{$^\mathrm{j}$}$20.50\pm 0.10^\mathrm{j}$ \\%UVES - H2
\hline
\end{tabular}
\flushleft
$^\mathrm{a}$ Rest-frame equivalent width of the absorption detected at the position of the \ion{C}{i} lines (measurements drawn from SDSS spectra).\\
$^\mathrm{b}$ Average signal-to-noise ratio per pixel around the \ion{C}{i} lines.\\
$^\mathrm{c}$ Neutral atomic-hydrogen column density of the systems measured from Ly$\alpha$ absorption in SDSS spectra.\\
$^\mathrm{d}$ $N(\ion{H}{i})$ values derived from follow-up UVES spectroscopy.\\
$^\mathrm{e}$ \citet{2009A&A...505.1087N};
$^\mathrm{f}$ \citet{2012ApJ...760...42W};
$^\mathrm{g}$ \citet{2010A&A...523A..80N};
$^\mathrm{h}$ \citet{2008A&A...491..397N};
$^\mathrm{i}$ \citet{2010MNRAS.408.2071M};
$^\mathrm{j}$ \citet{2006A&A...457...71L}.
\end{table*}

\addtocounter{table}{-1}
\begin{table*}[htb]
  \caption{{\it continued.}}
\begin{tabular}{ccccccccccc}
\hline
\hline
QSO name & plate & MJD & fibre &
$z_\mathrm{em}$  &
$z_\mathrm{abs}$ &
$W_\mathrm{r}(\lambda 1560)^\mathrm{a}$ &
$W_\mathrm{r}(\lambda 1656)^\mathrm{a}$ & S/N$^\mathrm{b}$ & $\log N(\ion{H}{i})^\mathrm{c}_\mathrm{SDSS}$ & $\log N(\ion{H}{i})^\mathrm{d}_\mathrm{UVES}$\\
& & & & & & [\AA ] & [\AA ] & & [atoms~cm$^{-2}$] & [atoms~cm$^{-2}$]\\
\hline
J\,222910.16$+$141402.2 & 0738 & 52521 & 382 & 2.11 & 1.586 & $0.33\pm 0.13$ & $0.47\pm 0.13$ & \phantom{1}6 &       ...               &       ...               \\%XS?
J\,225719.04$-$100104.7 & 0724 & 52254 & 150 & 2.08 & 1.836 & $0.64\pm 0.05$ & $0.89\pm 0.05$ &           17 &       ...               & $20.38\pm 0.07$         \\%UVES
J\,233133.05$-$090246.6 & 0646 & 52523 & 619 & 2.44 & 1.734 & $0.41\pm 0.04$ & $0.35\pm 0.04$ &           20 &       ...               &       ...               \\%
J\,233156.49$-$090802.0 & 0646 & 52523 & 616 & 2.66 & 2.143 & $1.34\pm 0.13$ & $1.17\pm 0.11$ & \phantom{1}8 &       ...               & $21.15\pm 0.15$         \\%UVES Nestor - CO
J\,233633.81$-$105841.5 & 0647 & 52553 & 201 & 2.04 & 1.829 & $0.22\pm 0.05$ & $0.40\pm 0.07$ &           15 &       ...               &       ...               \\%XS
J\,234023.67$-$005327.1 & 0385 & 51877 & 204 & 2.09 & 2.054 & $0.11\pm 0.02$ & $0.17\pm 0.02$ &           37 &       ...               & $20.33\pm 0.03$          \\%UVES HIRES - HD/H2
J\,235057.87$-$005210.0 & 0386 & 51788 & 137 & 3.03 & 2.426 & $0.62\pm 0.08$ & $0.63\pm 0.08$ &           11 & \phantom{$^\mathrm{e}$}$20.39\pm 0.27^\mathrm{e}$ & \phantom{$^\mathrm{j}$}$20.50\pm 0.10^\mathrm{j}$ \\%UVES - H2
\hline
\end{tabular}
\flushleft
$^\mathrm{a}$ Rest-frame equivalent width of the absorption detected at the position of the \ion{C}{i} lines (measurements drawn from SDSS spectra).\\
$^\mathrm{b}$ Average signal-to-noise ratio per pixel around the \ion{C}{i} lines.\\
$^\mathrm{c}$ Neutral atomic-hydrogen column density of the systems measured from Ly$\alpha$ absorption in SDSS spectra.\\
$^\mathrm{d}$ $N(\ion{H}{i})$ value derived from follow-up UVES spectroscopy.\\
$^\mathrm{e}$ \citet{2009A&A...505.1087N};
$^\mathrm{f}$ \citet{2012ApJ...760...42W};
$^\mathrm{g}$ \citet{2010A&A...523A..80N};
$^\mathrm{h}$ \citet{2008A&A...491..397N};
$^\mathrm{i}$ \citet{2010MNRAS.408.2071M};
$^\mathrm{j}$ \citet{2006A&A...457...71L}.
\end{table*}

%--------------------------------------------------------------------------------

\begin{table*}[htb]
  \caption{Intervening \ion{C}{i}-absorber number counts.\label{tab:dNdz}}
\begin{tabular}{lcllcccr}
\hline
\hline
\ion{C}{i} samples & $W_\mathrm{r}(\lambda 1560)$ & $z_\mathrm{min}$ & $z_\mathrm{max}$ & $\Delta z$ & $\#$ of \ion{C}{i} & $n_\mathrm{\ion{C}{i}}^\mathrm{a}$ \\
& [\AA] & & & & & ($\times 10^{-3}$) &\\
\hline
All systems                & $\ge 0.4$           & 1.5 & 1.9   &  6268 & 22 & $3.5\pm 0.8$ & \\
                                   &                           & 1.9 & 3.35 &  6710 &  8 & $1.2\pm 0.4$ & \\
Strongest                    & $\ge 0.64$         & 1.5 & 1.9   &  6268 & 10 & $1.6\pm 0.5$ & \\
                                   &                           & 1.9 & 3.35 &  6710 &  4 & $0.6\pm 0.3$ & \\
Weaker                       & $\ge 0.4,<0.64$ & 1.5 & 1.9   &  6268 & 12 & $1.9\pm 0.6$ & \\
                                   &                           & 1.9 & 3.35 &  6710 &  4 & $0.6\pm 0.3$ & \\
\hline
\end{tabular}
\flushleft
$^\mathrm{a}$ {\bf Prior to correcting for incompleteness (i.e., $\sim 18$\%
at $W_\mathrm{r}(\lambda 1560)=0.4$~\AA).}
%$^\mathrm{a}$ Excluding proximate systems.\\
%$^\mathrm{b}$ Median signal-to-noise ratio per pixel of the spectra in the wavelength ranges where \ion{C}{i} absorption has been searched for.
\end{table*}

%--------------------------------------------------------------------------------

\begin{table*}[htb]
  \caption{SED fitting of QSO spectra with detected \ion{C}{i} absorbers.\label{tab:sed}}
\begin{tabular}{cccclc|rccccc}
\hline
\hline
QSO name & $z_\mathrm{em}$ & $z_\mathrm{abs}$ & $E($B-V$)$ & best fit & $A_\mathrm{bump}$ & C.S.$^\mathrm{a}$: & \multicolumn{2}{c}{$E($B-V$)$ [mag]} & \multicolumn{2}{c}{$A_\mathrm{bump}$}\\
& & & [mag] & & & $N_\mathrm{QSO}$ & median & $\sigma$ & median & $\sigma$\\
\hline
J\,021606.13$-$002104.9 & 2.22 & 1.737 &    0.039 & SMC  &    0.097 & 173  & $-$0.008 & 0.053 & $-$0.009 & 0.123\\
J\,030049.24$-$072137.8 & 2.11 & 1.536 &    0.075 & SMC  &    0.160 & 448  & $-$0.004 & 0.048 &    0.002 & 0.136\\
J\,080801.74$+$330009.2 & 1.90 & 1.888 &    0.029 & SMC  & $-$0.032 & 151  &    0.001 & 0.044 & $-$0.028 & 0.112\\
J\,081116.06$+$083837.7 & 2.07 & 1.906 &    0.081 & SMC  & $-$0.027 & 247  & $-$0.003 & 0.049 & $-$0.031 & 0.127\\
J\,081540.60$+$264021.6 & 1.94 & 1.681 &    0.109 & LMC  &    0.364 & 428  & $-$0.012 & 0.043 & $-$0.027 & 0.142\\
J\,082003.40$+$155932.9 & 1.95 & 1.547 &    0.203 & LMC2 &    0.420 & 373  & $-$0.004 & 0.069 & $-$0.012 & 0.142\\
J\,085206.65$+$193548.4 & 1.91 & 1.788 &    0.050 & SMC  &    0.032 & 580  & $-$0.003 & 0.044 & $-$0.020 & 0.112\\
J\,085437.60$+$031734.8 & 2.24 & 1.567 &    0.070 & SMC  &    0.263 & 192  & $-$0.003 & 0.054 &    0.050 & 0.106\\
J\,085726.79$+$185524.3 & 1.89 & 1.730 &    0.016 & SMC  &    0.063 & 81   & $-$0.004 & 0.053 & $-$0.064 & 0.107\\
J\,090558.75$+$553039.2 & 1.57 & 1.579 & $-$0.019 & SMC  &    0.070 & 692  & $-$0.002 & 0.041 &    0.011 & 0.080\\
J\,090942.56$+$532126.5 & 2.07 & 1.628 &    0.003 & SMC  &    0.003 & 527  & $-$0.005 & 0.044 & $-$0.019 & 0.140\\
J\,091516.27$+$071750.9 & 2.38 & 1.681 & $-$0.010 & SMC  &    0.117 & 40   & $-$0.017 & 0.077 &    0.019 & 0.206\\
J\,091721.37$+$015448.1 & 2.18 & 2.107 &    0.121 & SMC  &    0.227 & 86   & $-$0.014 & 0.031 &    0.006 & 0.114\\
J\,092759.79$+$154321.8 & 1.81 & 1.731 &    0.217 & LMC2 &    0.331 & 357  & $-$0.008 & 0.059 & $-$0.018 & 0.092\\
J\,094529.37$+$552525.7 & 2.24 & 1.867 &    0.120 & SMC  &    0.015 & 314  & $-$0.006 & 0.043 & $-$0.016 & 0.130\\
J\,095114.97$+$545736.5 & 1.80 & 1.613 & $-$0.004 & SMC  & $-$0.027 & 397  & $-$0.004 & 0.042 & $-$0.015 & 0.109\\
J\,101544.34$+$310617.2 & 1.56 & 1.596 &    0.016 & SMC  & $-$0.006 & 685  &    0.000 & 0.043 &    0.012 & 0.085\\
J\,104705.76$+$205734.5 & 2.01 & 1.775 &    0.159 & LMC2 &    0.277 & 282  & $-$0.021 & 0.055 & $-$0.036 & 0.133\\
J\,105436.96$+$542343.7 & 1.50 & 1.513 &    0.014 & SMC  & $-$0.008 & 319  &    0.001 & 0.037 &    0.007 & 0.072\\
J\,105746.42$+$662250.5 & 2.34 & 2.088 &    0.040 & SMC  &    0.112 & 277  & $-$0.010 & 0.037 & $-$0.046 & 0.120\\
J\,111756.53$+$143716.9 & 2.09 & 2.001 &    0.049 & SMC  &    0.085 & 560  & $-$0.005 & 0.036 & $-$0.030 & 0.097\\
J\,112250.95$+$143732.5 & 2.05 & 1.554 & $-$0.028 & SMC  &    0.032 & 106  &    0.003 & 0.047 & $-$0.026 & 0.138\\
J\,112913.29$-$023740.9 & 1.86 & 1.623 &    0.020 & SMC  & $-$0.151 & 603  & $-$0.003 & 0.045 & $-$0.026 & 0.124\\
J\,113341.29$-$005740.1 & 1.68 & 1.706 &    0.266 & SMC  &    0.062 & 386  & $-$0.002 & 0.042 &    0.002 & 0.094\\
J\,114141.84$+$444206.1 & 1.96 & 1.903 &    0.149 & LMC2 &    0.368 & 398  & $-$0.023 & 0.060 & $-$0.028 & 0.109\\
J\,115153.87$+$150945.0 & 3.05 & 2.400 & $-$0.007 & SMC  &    0.110 & 92   & $-$0.007 & 0.033 & $-$0.105 & 0.176\\
J\,115705.51$+$615521.7 & 2.51 & 2.460 &    0.215 & SMC  &    0.370 & 148  & $-$0.014 & 0.033 & $-$0.087 & 0.101\\
J\,120935.79$+$671715.7 & 2.03 & 1.843 &    0.094 & MW   &    0.394 & 300  & $-$0.013 & 0.028 & $-$0.039 & 0.119\\
J\,122825.67$+$303038.6 & 3.91 & 3.098 & $-$0.034 & SMC  &    0.067 & 43   & $-$0.008 & 0.061 &    0.412 & 0.471\\
J\,123714.61$+$064759.6 & 2.78 & 2.691 &    0.143 & LMC2 &    0.431 & 35   & $-$0.026 & 0.074 & $-$0.168 & 0.156\\
J\,124708.43$+$500320.8 & 2.27 & 2.135 &    0.075 & LMC  &    0.319 & 357  & $-$0.012 & 0.029 & $-$0.037 & 0.124\\
J\,124829.71$+$284858.1 & 1.54 & 1.513 & $-$0.035 & SMC  & $-$0.049 & 1035 & $-$0.002 & 0.041 &    0.022 & 0.092\\
J\,124841.45$+$302433.0 & 2.06 & 1.691 & $-$0.025 & SMC  & $-$0.019 & 583  & $-$0.004 & 0.042 & $-$0.024 & 0.136\\
J\,125552.60$+$223424.4 & 1.82 & 1.526 &    0.036 & SMC  &    0.141 & 898  & $-$0.004 & 0.041 & $-$0.015 & 0.134\\
J\,130225.28$+$211158.6 & 1.76 & 1.656 & $-$0.004 & SMC  & $-$0.139 & 452  & $-$0.012 & 0.036 & $-$0.002 & 0.114\\
J\,130628.87$+$281550.8 & 2.10 & 2.012 &    0.078 & SMC  &    0.015 & 465  & $-$0.008 & 0.034 & $-$0.030 & 0.099\\
J\,130828.43$+$584000.6 & 3.09 & 2.473 &    0.020 & SMC  & $-$0.158 & 54   & $-$0.010 & 0.037 & $-$0.144 & 0.211\\
J\,131129.11$+$222552.6 & 3.14 & 3.093 &    0.036 & SMC  &    0.067 & 82   & $-$0.011 & 0.036 & $-$0.054 & 0.345\\
J\,131400.57$+$054319.5 & 1.89 & 1.583 &    0.046 & SMC  &    0.270 & 340  & $-$0.006 & 0.044 & $-$0.013 & 0.120\\
J\,134122.51$+$185214.0 & 2.00 & 1.544 & $-$0.016 & SMC  &    0.000 & 32   &    0.003 & 0.053 & $-$0.001 & 0.185\\
J\,134601.10$+$064408.4 & 2.09 & 1.512 &    0.015 & SMC  &    0.034 & 118  & $-$0.015 & 0.040 &    0.036 & 0.130\\
J\,135122.00$+$461828.5 & 1.81 & 1.606 &    0.142 & LMC2 &    0.408 & 680  & $-$0.005 & 0.067 & $-$0.027 & 0.123\\
J\,141550.47$+$300146.9 & 2.08 & 1.676 &    0.074 & SMC  &    0.226 & 83   & $-$0.001 & 0.045 & $-$0.033 & 0.245\\
J\,141606.79$+$180403.2 & 2.13 & 1.622 &    0.056 & SMC  & $-$0.027 & 108  & $-$0.016 & 0.035 &    0.018 & 0.150\\
J\,143243.93$+$330746.7 & 2.09 & 2.058 &    0.010 & SMC  & $-$0.005 & 352  & $-$0.003 & 0.041 & $-$0.039 & 0.095\\
J\,143657.87$+$291100.6 & 1.77 & 1.596 &    0.147 & LMC  &    0.514 & 812  & $-$0.003 & 0.049 & $-$0.014 & 0.111\\
J\,143912.05$+$111740.6 & 2.58 & 2.418 &    0.172 & LMC2 &    0.271 & 52   & $-$0.033 & 0.145 & $-$0.080 & 0.098\\
J\,144929.27$+$333811.0 & 2.17 & 2.021 &    0.018 & SMC  & $-$0.157 & 89   & $-$0.009 & 0.029 &    0.025 & 0.137\\
J\,145432.54$+$343523.9 & 1.61 & 1.580 &    0.293 & LMC2 &    0.631 & 346  & $-$0.002 & 0.058 & $-$0.002 & 0.089\\
J\,145953.25$+$012944.2 & 1.66 & 1.623 &    0.157 & LMC2 &    0.331 & 842  & $-$0.006 & 0.057 &    0.011 & 0.089\\
J\,150738.73$+$415530.6 & 1.79 & 1.674 &    0.056 & LMC  &    0.251 & 806  & $-$0.005 & 0.040 & $-$0.003 & 0.111\\
J\,152209.12$+$083020.0 & 1.93 & 1.627 &    0.017 & SMC  & $-$0.057 & 169  & $-$0.016 & 0.035 & $-$0.031 & 0.134\\
J\,160320.76$+$170117.8 & 1.99 & 1.890 & $-$0.010 & SMC  &    0.030 & 406  & $-$0.004 & 0.043 & $-$0.033 & 0.108\\
J\,160457.51$+$220300.5 & 1.98 & 1.641 &    0.263 & LMC2 &    0.735 & 225  & $-$0.007 & 0.075 & $-$0.041 & 0.151\\
J\,161526.65$+$264813.8 & 2.18 & 2.118 &    0.035 & SMC  &    0.120 & 324  & $-$0.005 & 0.042 & $-$0.037 & 0.095\\
J\,162321.43$+$135532.4 & 1.75 & 1.751 &    0.031 & SMC  &    0.034 & 816  & $-$0.005 & 0.036 &    0.006 & 0.093\\
J\,164610.20$+$232923.0 & 2.06 & 1.998 & $-$0.027 & SMC  & $-$0.031 & 284  & $-$0.004 & 0.043 & $-$0.037 & 0.092\\
J\,170542.92$+$354340.4 & 2.01 & 2.038 &    0.146 & LMC2 &    0.364 & 210  & $-$0.009 & 0.075 & $-$0.030 & 0.077\\
J\,212329.47$-$005053.0 & 2.26 & 2.060 & $-$0.011 & SMC  &    0.000 & 32   &    0.002 & 0.041 & $-$0.023 & 0.123\\
J\,222910.16$+$141402.2 & 2.11 & 1.586 & $-$0.024 & SMC  &    0.026 & 538  & $-$0.004 & 0.041 &    0.006 & 0.136\\
J\,225719.04$-$100104.7 & 2.08 & 1.836 &    0.045 & SMC  &    0.177 & 192  & $-$0.007 & 0.039 & $-$0.040 & 0.129\\
J\,233133.05$-$090246.6 & 2.44 & 1.734 &    0.101 & LMC  &    0.393 & 38   &    0.002 & 0.089 &    0.084 & 0.215\\
J\,233156.49$-$090802.0 & 2.66 & 2.143 &    0.136 & SMC  &    0.234 & 46   & $-$0.008 & 0.078 & $-$0.015 & 0.104\\
J\,233633.81$-$105841.5 & 2.04 & 1.829 &    0.004 & SMC  & $-$0.103 & 484  & $-$0.003 & 0.045 & $-$0.026 & 0.123\\
J\,234023.67$-$005327.1 & 2.09 & 2.054 &    0.057 & SMC  & $-$0.023 & 34   & $-$0.001 & 0.047 & $-$0.071 & 0.113\\
J\,235057.87$-$005210.0 & 3.03 & 2.426 &    0.003 & SMC  &    0.001 & 42   & $-$0.013 & 0.067 & $-$0.098 & 0.169\\
\hline
\end{tabular}
\flushleft
$^\mathrm{a}$ C.S. $=$ control sample.
\end{table*}

%--------------------------------------------------------------------------------

%------------------------------------------------------------------------------

\end{document}